\documentclass[amsmath,amssymb,twocolumn,showpacs]{revtex4}

\usepackage{graphicx}
\usepackage{subfigure}
\usepackage[usenames]{color}

\begin{document}

\title{Anisotropic in-plane optical conductivity in detwinned Ba(Fe$_{1-x}$Co$_x$)$_2$As$_2$}
\author{A. Dusza$^{1*}$, A. Lucarelli$^{1*}$, A. Sanna$^2$, S. Massidda$^2$, J.-H. Chu$^3$, I.R. Fisher$^3$ and L. Degiorgi$^1$} \affiliation{$^1$Laboratorium f\"ur Festk\"orperphysik, ETH - Z\"urich, CH-8093 Z\"urich, Switzerland}
\affiliation{$^2$Dipartimento di Fisica, Universit\'a degli Studi di Cagliari, IT-09042 Monserrato, Italy}
\affiliation{$^3$Geballe Laboratory for Advanced Materials and Department of Applied Physics, Stanford University, Stanford, California
94305-4045, USA and Stanford Institute for Materials and Energy Sciences, SLAC National Accelerator Laboratory, 2575 Sand Hill Road, Menlo Park, California 94025, U.S.A.}

\date{\today}

\begin{abstract}
We study the anisotropic in-plane optical conductivity of detwinned Ba(Fe$_{1-x}$Co$_x$)$_2$As$_2$ single crystals for $x$=0, 2.5$\%$ and 4.5$\%$ in a broad energy range (3 meV-5 eV) across their structural and magnetic transitions. For temperatures below the Neel transition, the topology of the reconstructed Fermi surface, combined with the distinct behavior of the scattering rates, determines the anisotropy of the low frequency optical response. For the itinerant charge carriers, we are able to disentangle the evolution of the Drude weights and scattering rates and to observe their enhancement along the orthorhombic antiferromagnetic $a$-axis with respect to the ferromagnetic $b$-axis. For temperatures above $T_s$, uniaxial stress leads to a finite in-plane anisotropy. The anisotropy of the optical conductivity, leading to a significant dichroism, extends to high frequencies in the mid- and near-infrared regions. The temperature dependence of the dichroism at all dopings scales with the anisotropy ratio of the $dc$ conductivity, suggesting the electronic nature of the structural transition. Our findings bear testimony to a large nematic susceptibility that couples very effectively to the uniaxial lattice strain. In order to clarify the subtle interplay of magnetism and Fermi surface topology we compare our results with theoretical calculations obtained from density functional theory within the full-potential linear augmented plane-wave method.
\end{abstract}



\maketitle

\section{Introduction}
In many unconventional superconductors including cuprates and Fe-based pnictides, superconductivity emerges from a complicated soup of competing phases in the normal state when magnetism is suppressed by doping, pressure or other external parameters \cite{norman,basov_review}. This multiplicity of phases includes nematicity, defined as the spontaneously broken C4 rotational
symmetry of the square lattice, and a novel form of magnetism, arising from either orbital currents or antiferromagnetic fluctuations. In the Fe-based pnictide superconductors (for a review with a comprehensive references list see Ref. \onlinecite{johnson}), nematic correlations and antiferromagnetic fluctuations have also been recently connected with symmetry breaking competing phases. Inelastic neutron scattering experiments revealed anisotropic magnetic excitations \cite{li}, coupled to the structural tetragonal-orthorhombic transition at $T_s$. This structural transition elongates the Fe-Fe distance in the $ab$ plane along the $a$-axis direction and contracts it along the perpendicular $b$-axis. It turns out that below the Neel temperature $T_N$ ($\le T_s$) the spins present ferromagnetic correlations along the shorter orthorhombic $b$-axis and antiferromagnetic stripes along the longer $a$-axis \cite{li}. The two-fold anisotropy is also evident in quasiparticle interference observed via scanning tunnelling microscopy measurements \cite{chuang} and further confirmed by angle-resolved-photoemission-spectroscopy (ARPES) data, collected on crystals for which the incident beam size was comparable to the size of a single structural domain \cite{wang}. Furthermore, quantum oscillations in the parent compound reveal that the reconstructed Fermi-surface (FS) comprises several small pockets \cite{Terashima}. The smallest of these pockets is essentially isotropic in the $ab$-plane, but the other, larger pockets are much more anisotropic. 

The first ARPES observations of an anisotropic electronic dispersion \cite{wang} motivated an intensive research activity also with probes for which any impact of the electronic anisotropy would be obscured by the formation of dense twin domains. These correspond to adjacent microscopic domains as small as a few microns with alternating orthorhombic $a$ and $b$ axes \cite{tanatar}. Two distinct methods have been employed so far to detwin the specimen: application of uniaxial stress \cite{chudw,tanatar} and of an in-plane magnetic field \cite{chu_magnetic}. The former method, employed here, is superior in order to achieve an almost complete detwinning. Recent ARPES measurements \cite{zxshen,kim} of detwinned single crystals of Ba(Fe$_{1-x}$Co$_x$)$_2$As$_2$ reveal an increase (decrease) in the binding energy of bands with dominant $d_{yz}$ ($d_{xz}$) character on cooling through $T_s$ \cite{zxshen}, leading to a difference in orbital occupancy. The splitting of the $d_{xz}$ and $d_{yz}$ bands is progressively diminished with Co substitution in Ba(Fe$_{1-x}$Co$_x$)$_2$As$_2$, reflecting the monotonic decrease in the lattice orthorhombicity $\lbrack 2(a-b)/(a+b)\rbrack$. For temperatures above $T_s$, the band-splitting can be induced up to rather high temperatures by uniaxial stress. 

Mechanically detwinned crystals also provide a suitable playground in order to explore the intrinsic in-plane anisotropy of the transport properties. Measurements of the $dc$ resistivity as a function of temperature of the single domain parent compounds BaFe$_2$As$_2$, SrFe$_2$As$_2$ and CaFe$_2$As$_2$ (i.e., so called 122 iron pnictides) reveal a modest in-plane $dc$ anisotropy for temperatures below $T_s$, with the resistivity in the ferromagnetic direction larger than along the antiferromagnetic direction \cite{chudw,Tanatar, Blomberg}. Substitution of Co, Ni or Cu suppresses the lattice orthorhombicity \cite{Prozorov}, but in contrast the in-plane resistivity anisotropy is found to initially increase with the concentration of the substituent, before reverting to an isotropic in-plane conductivity once the structural transition is completely suppressed \cite{chudw,Kuo}. Perhaps coincidentally, the onset of the large in-plane anisotropy for the cases of Co and Ni substitution occurs rather abruptly at a composition close to the start of the superconducting dome. 

For temperatures above $T_s$, there is a remarkably large sensitivity to uniaxial pressure, leading to a large induced in-plane resistivity anisotropy that is not observed for overdoped compositions \cite{chudw}. There is no evidence in thermodynamic or transport measurements for an additional phase transition above $T_s$ for unstressed crystals, implying that the induced anisotropy is the result of a large nematic susceptibility, rather than the presence of static nematic order. The observation of a large in-plane resistivity anisotropy, at least for the electron-doped Ò122Ó Fe arsenides, bears witness to the orthorhombicity of the material, but does not distinguish between anisotropy in the electronic structure and anisotropy in the scattering rate. To this end, reflectivity measurements of detwinned single crystals using polarized light can provide important insight to the effects of the magnetic and structural transitions on the anisotropic charge dynamics and the electronic band structure. Indeed, the counterintuitive anisotropic behavior of $\rho(T)$ is also reflected in the finite frequency response of the charge carriers as observed by the optical measurements reported in our previous Letter work \cite{dusza}. Optical measurements of detwinned single crystals of Ba(Fe$_{1-x}$Co$_x$)$_2$As$_2$ in the underdoped regime reveal large changes in the low-frequency metallic response on cooling through $T_s$ and $T_N$ together with a pronounced optical anisotropy (i.e.,  $\Delta\sigma_1(\omega)=\sigma_1(\omega,E\parallel a)-\sigma_1(\omega,E\parallel b)$) at high frequencies, defining the linear dichroism \cite{dusza}. For light polarized in the antiferromagnetic $a$ direction, there is an increase in the scattering rate, but this is accompanied by a dramatic increase in the metallic spectral weight that ultimately leads to a reduction in the $dc$ resistivity, consistent with observations. For light polarized along the $b$ direction, the dominant effect is a reduction in the metallic spectral weight, consistent with the increase in the $dc$ resistivity. The high frequency dichroism, which is smaller for higher Co concentrations, clearly reveals that changes in the electronic structure are not confined to near the Fermi energy. Similar to $dc$ transport and ARPES measurements, a pronounced optical anisotropy persists at temperatures above $T_s$ for crystals held under uniaxial stress, also anticipating a substantial nematic susceptibility.

In the present work, we expand in greater detail our broad spectral range characterization of the anisotropic optical conductivity of detwinned Ba(Fe$_{1-x}$Co$_x$)$_2$As$_2$ single crystals under uniaxial pressure. We extensively study the underdoped region at compositions with $x = 0$, $x = 0.025$ and $x = 0.045$ with light polarized along the in-plane orthorhombic $a$ and $b$ axes. We present here the full set of data and their complete phenomenological analysis. Furthermore, in order to clarify the subtle interplay of magnetism and Fermi surface topology we compare directly our optical measurements with theoretical calculations obtained from density functional theory within the full-potential linear augmented plane-wave method (LAPW) \cite{sanna}. 

\begin{figure}[!tb]
\center
\includegraphics[width=6cm]{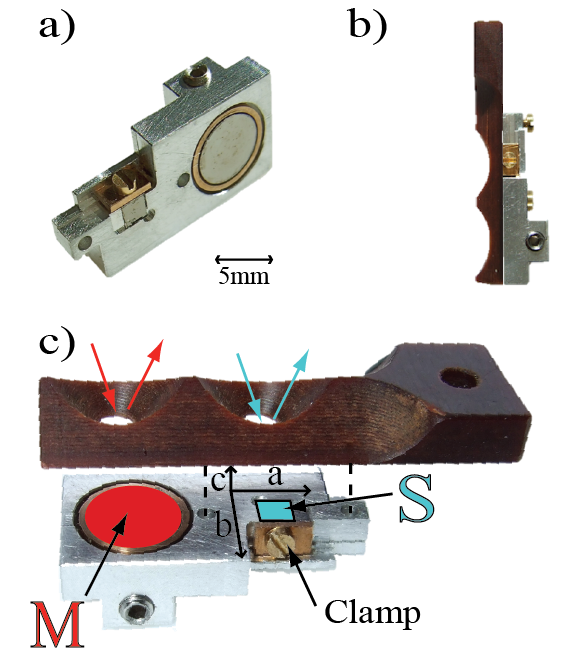}
\caption{(color online) Sample-holder set-up  including a mechanical clamp (a) and an optical mask (b and c). Uniaxial-pressure is applied parallel to the $b$-axis of the sample (S) by drawing the clamp against the side-surface of the crystal (a and c). The optical mask, attached in tight contact with the clamp device, shapes identically incident and reflected light beams for the tungsten reference mirror M (red arrows) and S (light blue arrows).} \label{Clamp}
\end{figure}

\section{Experimental}

\subsection{Samples}
Single crystals of Ba(Fe$_{1-x}$Co$_x$)$_2$As$_2$ with $x = 0$, 2.5 and 4.5\% were grown using a self-flux method \cite{chudw}. The crystals have a plate-like morphology of 0.1$\div$0.3 mm thickness, with the $c$-axis perpendicular to
the plane of the plates. Crystals were cut in to a square shape, approximately 2 mm on the side, oriented such that below $T_s$ the orthorhombic $a/b$ axes are parallel to the sides of the square \cite{chudw}. Detailed thermodynamic, transport and neutron scattering measurements for the studied dopings of Ba(Fe$_{1-x}$Co$_x$)$_2$As$_2$ give evidence for structural, magnetic and superconducting phase transitions occurring at different temperatures \cite{chu,lester}: for $x = 0$, the coincident structural (tetragonal-orthorhombic) and magnetic transitions where the system forms antiferromagnetically ordered stripes occur at $T_s$ = $T_N$ = 135 K, whereas for $x = 0.025$ they develop at $T_s$ = 98 K and $T_N$ = 92 K, respectively. The compound with $x = 0.045$ undergoes first a structural transition at $T_s$ = 66 K then a magnetic transition at $T_N$ = 58 K and finally a superconducting one at $T_c$ = 15 K. 

\subsection{Technique}
It has recently been shown that almost single-domain specimens can be achieved by application of uniaxial pressure in situ \cite{chudw}. This is crucial in order to reveal the intrinsic anisotropy of the orthorhombic phase. To this goal we have extended the basic cantilever concept originally developed for transport measurements to allow optical measurements under constant uniaxial pressure. The device consists of a mechanical clamp (Fig. \ref{Clamp}(a)) and an optical mask (Fig. \ref{Clamp}(b) and \ref{Clamp}(c)) attached on top of it in tight contact. The pressure-device was designed according to the following specific criteria:
i) the uniaxial stress is applied to the sample (S) by tightening a screw and drawing the clamp against the side of the crystal (Fig. \ref{Clamp}(a) and Fig. \ref{Clamp}(c)). Even though our clamp set-up still lacks of a precisely tunable pressure, the uniaxial stress was gradually increased, so to observe optical anisotropy. The applied pressure is modest, such that $T_{N}$ is unaffected, and can be adjusted over a limited range (up to approximately 5 MPa \cite{chudw}). Cooling samples under such uniaxial stress results in a significantly larger population of domains for which the shorter ferromagnetic $b$-axis is oriented along the direction of the applied stress, almost fully detwinning the crystals.
ii) The major axis of the tightening screw lies nearby and parallel to the surface of the sample so that the shear- and thermal-stress effects are minimized. The thermal expansion $\Delta$$L$ of the tightening screw, exerting the uniaxial pressure, can be estimated to be of the order of $\Delta$$L = \alpha$$LdT = 20 \mu$m (for screw-length $L = 5$ mm, typical metallic thermal expansion coefficient $\alpha = 2\cdot10^{-5}$K$^{-1}$, and thermal excursion $dT  = 200$ K). This corresponds to a relative variation of about 0.4\%. By reasonably assuming $\Delta$$L/L = \Delta$$p/p$, the influence of thermal expansion effects is then negligible.
iii) The clamp set-up leaves the (001) facet of the single domain samples exposed, enabling us to perform optical reflectivity measurements ($R$($\omega$)).
iv) The optical mask guarantees data collection on surfaces of the same dimension for S and reference mirror (M) and therefore on equivalent flat spots.

The reflectivity ($R(\omega)$) at room temperature was first collected from different spectrometers such as the Bruker IFS48 for the mid-infrared (MIR, 500-4000 cm$^{-1}$) and near-infrared (NIR, 4000-7000 cm$^{-1}$) measurements and a PerkinElmer Lambda 950 capable to measure absolute reflectivity from NIR up to the ultra-violet (UV) range, i.e. 3200-4.8x10$^4$ cm$^{-1}$. The detwinning device was then placed inside our cryostat and finely aligned with micrometric precision within the optical path of the Fourier transform infrared Bruker Vertex 80v interferometer, so that we could perform optical measurements of ($R(\omega)$) at different temperatures in the spectral range from the far-infrared (FIR, $\omega<400$ cm$^{-1}$) up to the MIR, i.e. between 30 and 6000 cm$^{-1}$. Light in all spectrometers was polarized along the $a$ and $b$ axes of the detwinned samples, thus giving access to the anisotropic optical functions. 
\begin{figure}[!t]
\center
\includegraphics[width=6cm]{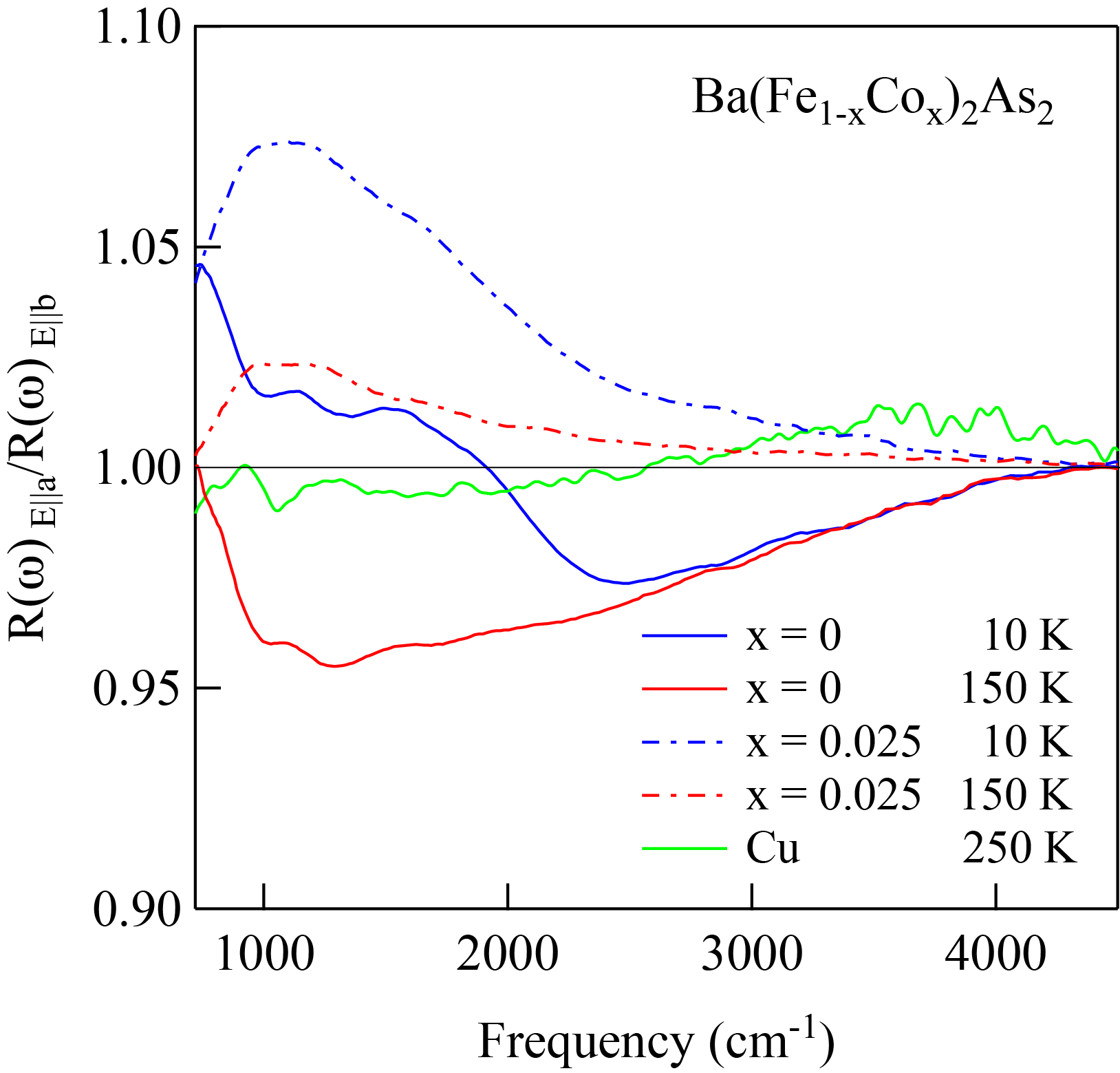}
\caption{(color online) Reflectivity ratio $R(\omega)_{E\parallel a}/R(\omega)_{E\parallel b}$ measured above and below $T_s$ along the $a$- and $b$-axis polarization directions for Ba(Fe$_{1-x}$Co$_x$)$_2$As$_2$ with $x = 0$ and 0.025 compared with the same ratio for a Cu sample of equivalent surface dimensions and thickness.} \label{CuRatio}
\end{figure}
\begin{figure*}[ht]
\center
\subfigure[]{\includegraphics[width=5.86cm]{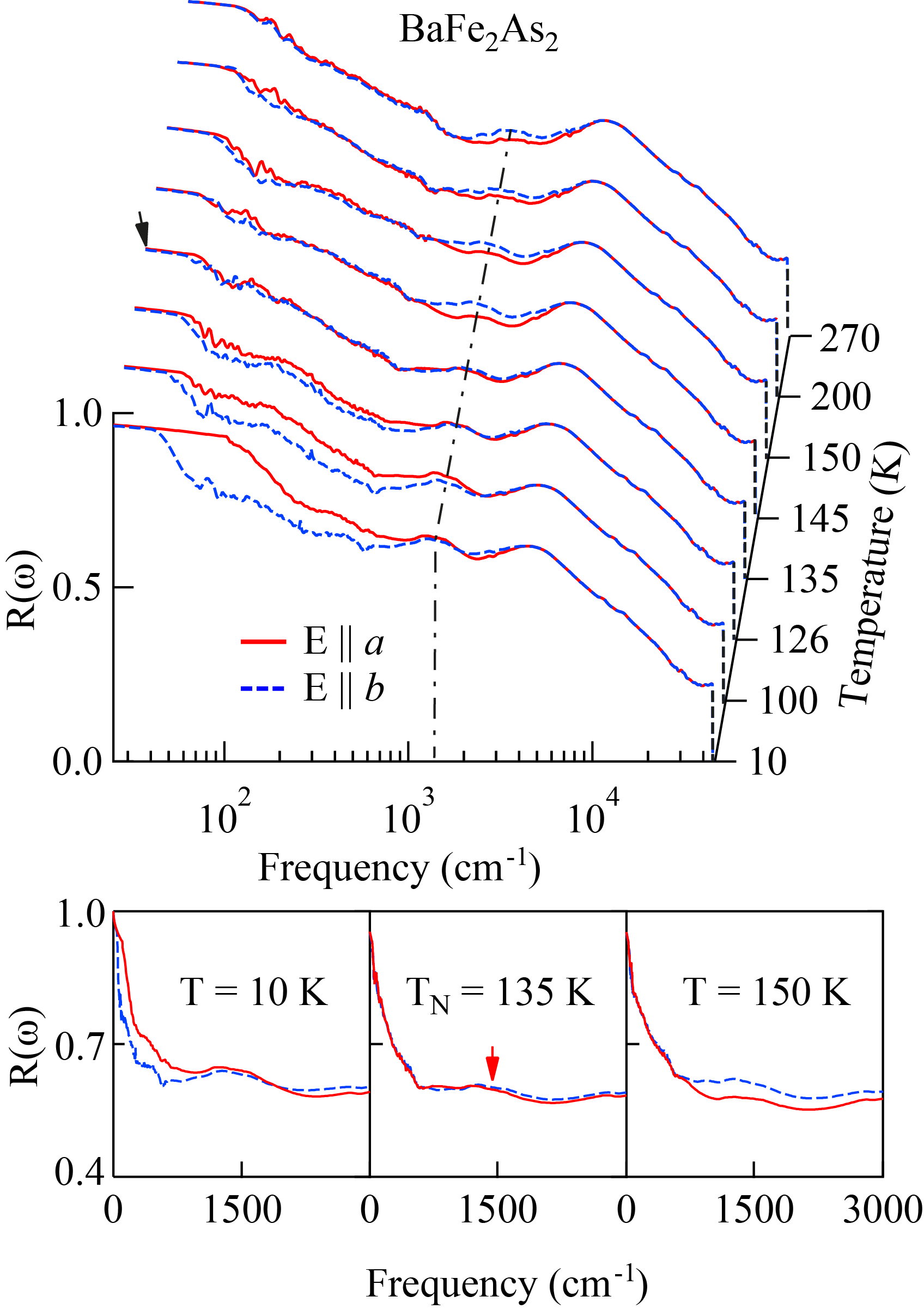}}
\subfigure[]{\includegraphics[width=5.86cm]{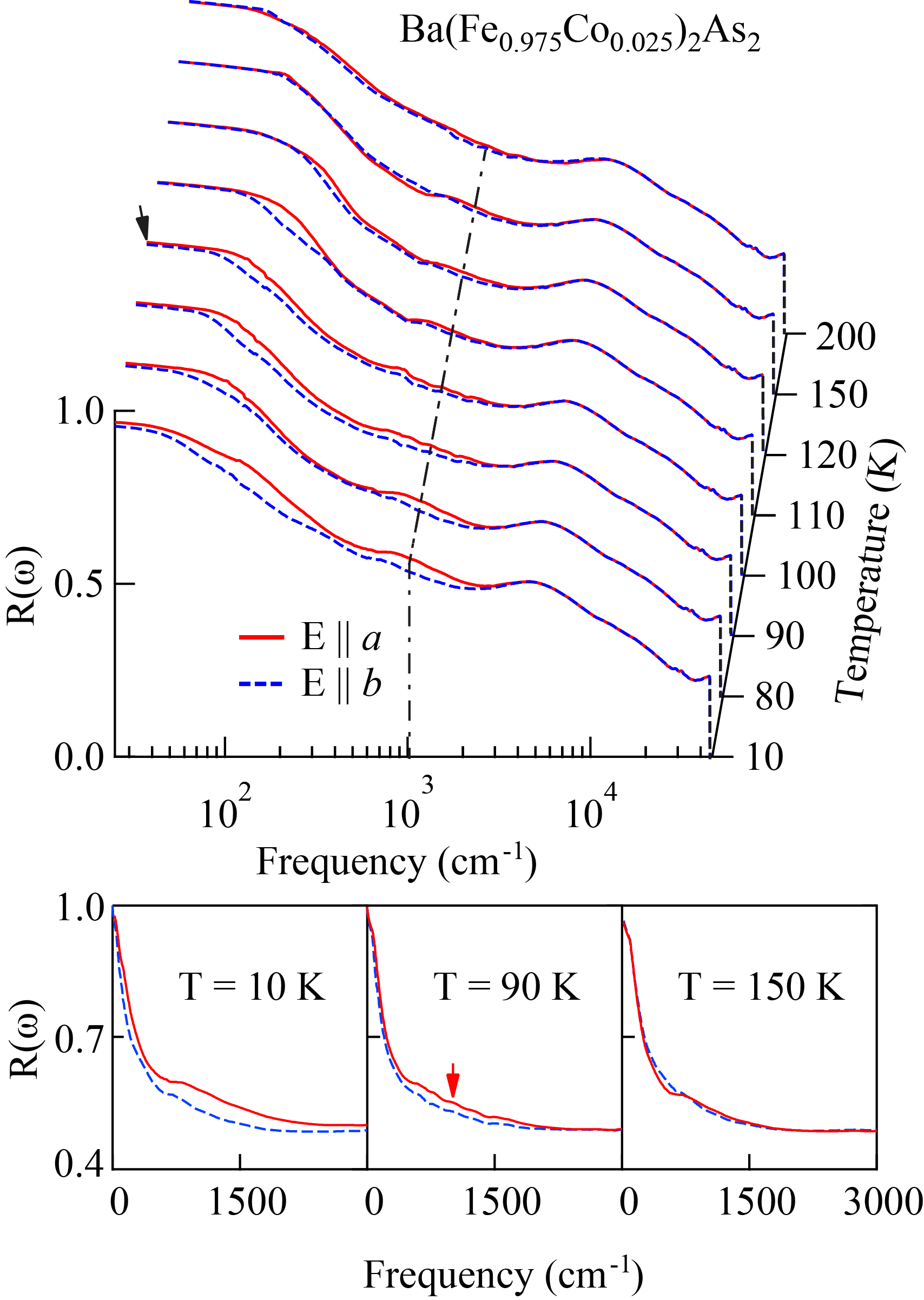}}
\subfigure[]{\includegraphics[width=5.86cm]{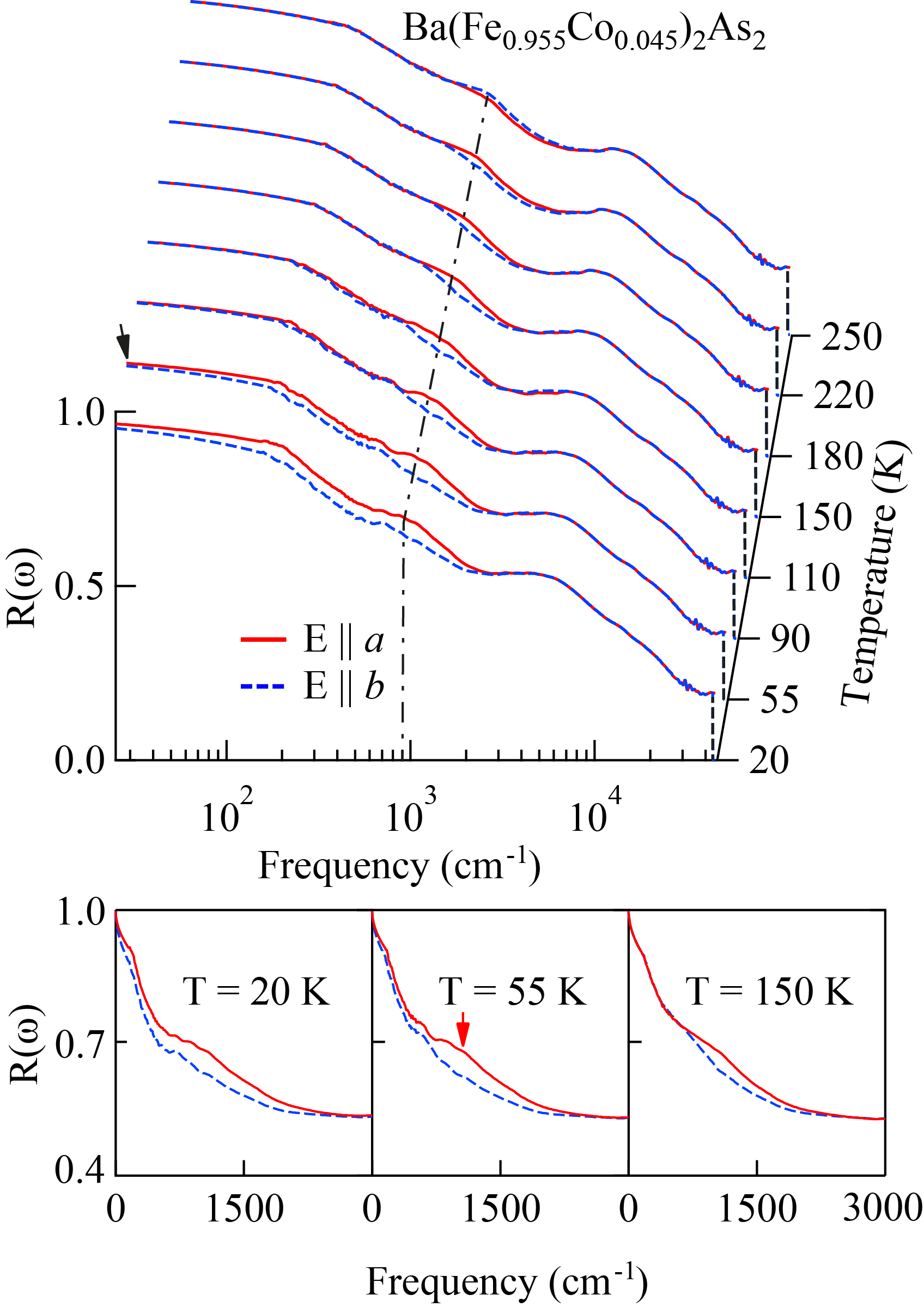}}
\caption{(color online) Temperature dependence of the optical reflectivity of detwinned BaFe$_2$As$_2$, Ba(Fe$_{0.975}$Co$_{0.025}$)$_2$As$_2$ and Ba(Fe$_{0.955}$Co$_{0.045}$)$_2$As$_2$ in the whole measured spectral range for two polarizations, parallel to the $a$-axis direction (E$\parallel$$a$, red solid line) or perpendicular to it (E$\parallel$$b$, blue dashed line). The black arrows indicate $R(\omega)$ at the temperature close to the structural phase transition at $T_s$. The dashed-dotted line (top panels) and the red arrows (bottom panels) indicate the center of the MIR  band at all temperatures and close to $T_N$, respectively.} \label{Ref}
\end{figure*}
The real part $\sigma$$_{1}(\omega)$ of the optical conductivity was obtained via the Kramers-Kronig transformation of $R(\omega)$ by applying suitable extrapolations at low and high frequencies. For the $\omega\to0$ extrapolation, we made use of the Hagen-Rubens (HR) formula ($R(\omega)=1-2\sqrt{\frac{\omega}{\sigma_{dc}}}$), inserting the $dc$ conductivity values ($\sigma_{dc}$) from Ref. \cite{chudw}, while above the upper frequency limit $R(\omega)\sim\omega^{-s}$ $(2\leqslant s \leqslant 4)$ \cite{grunerbook}. 

Several precautions were taken in order to avoid experimental artifacts:  i) The polarizers chosen for each measured frequency range have an extinction ratio greater than 200, thus reducing leakages below our 1\% error limit. ii) As control measurements for the detwinning setup we collected at different temperatures the optical reflectivity of a Cu sample of comparable surface dimensions and thickness with respect to the pnictide crystals and under equivalent uniaxial pressure. As expected, we could not observe any polarization dependence of the Cu reflectivity from room temperature down to 10 K (see e.g. the data at 250 K in Fig. \ref{CuRatio}). The Cu test measurements set to about 1-2\% the higher limit of the polarization dependence due to any possible experimental artifacts (i.e. bended surfaces, leakage of the polarizers etc.), which is notably lower then the anisotropy ratio measured for the iron-pnictides (Fig. 2 and 3). iii) Prior to performing optical experiments as a function of the polarization of light, the electrodynamic response of the twinned (i.e., unstressed) samples was first checked with unpolarized light, consistently recovering the same spectra previously presented in Ref. \cite{lucarelli}. vi) We achieved the same alignment conditions of M and S (Fig. \ref{Clamp}(c)) by imaging on both spots a red laser point source.
\begin{figure*}[ht]
\centering
\subfigure[]{\includegraphics[width=5.91cm]{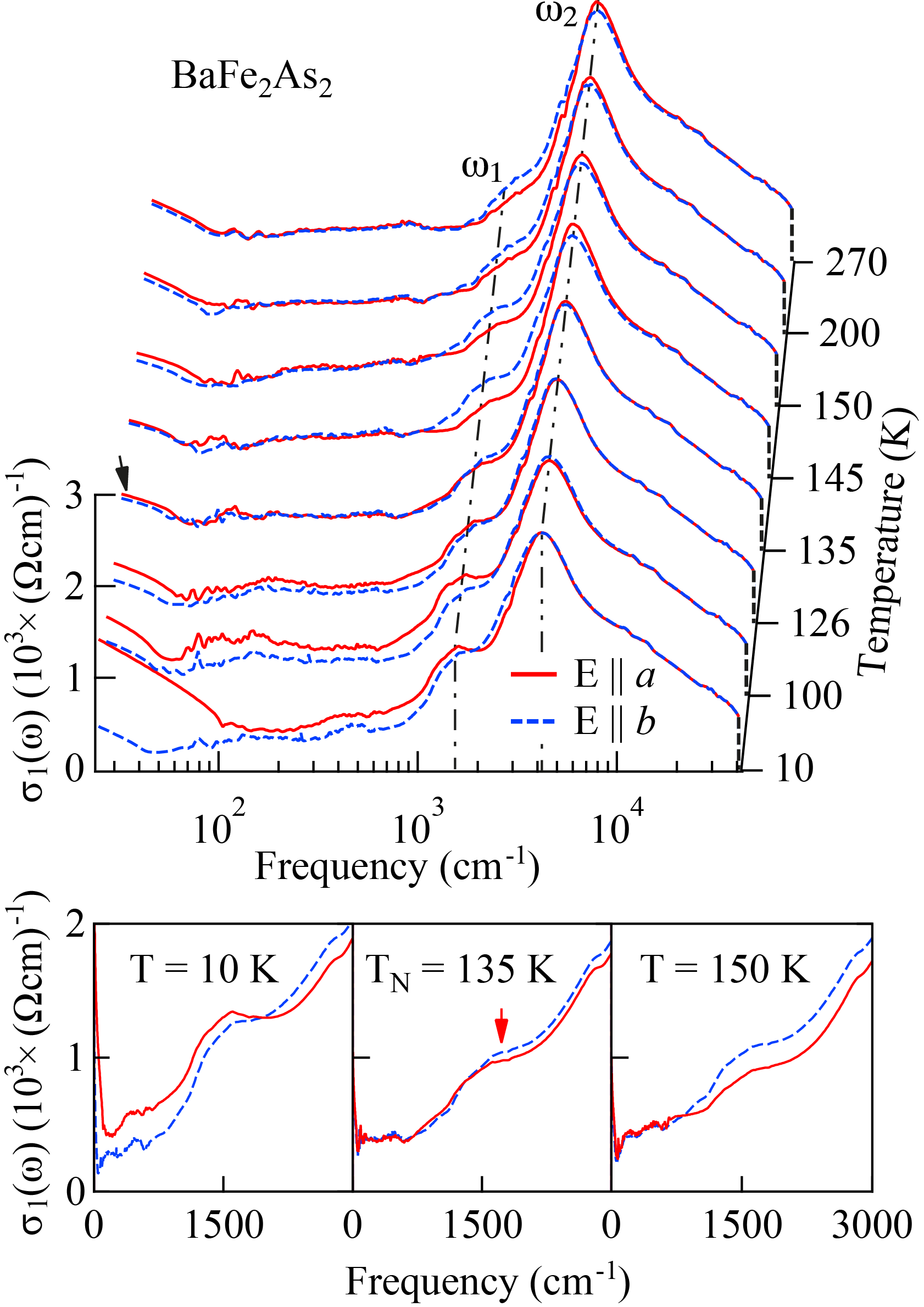}}
\subfigure[]{\includegraphics[width=5.91cm]{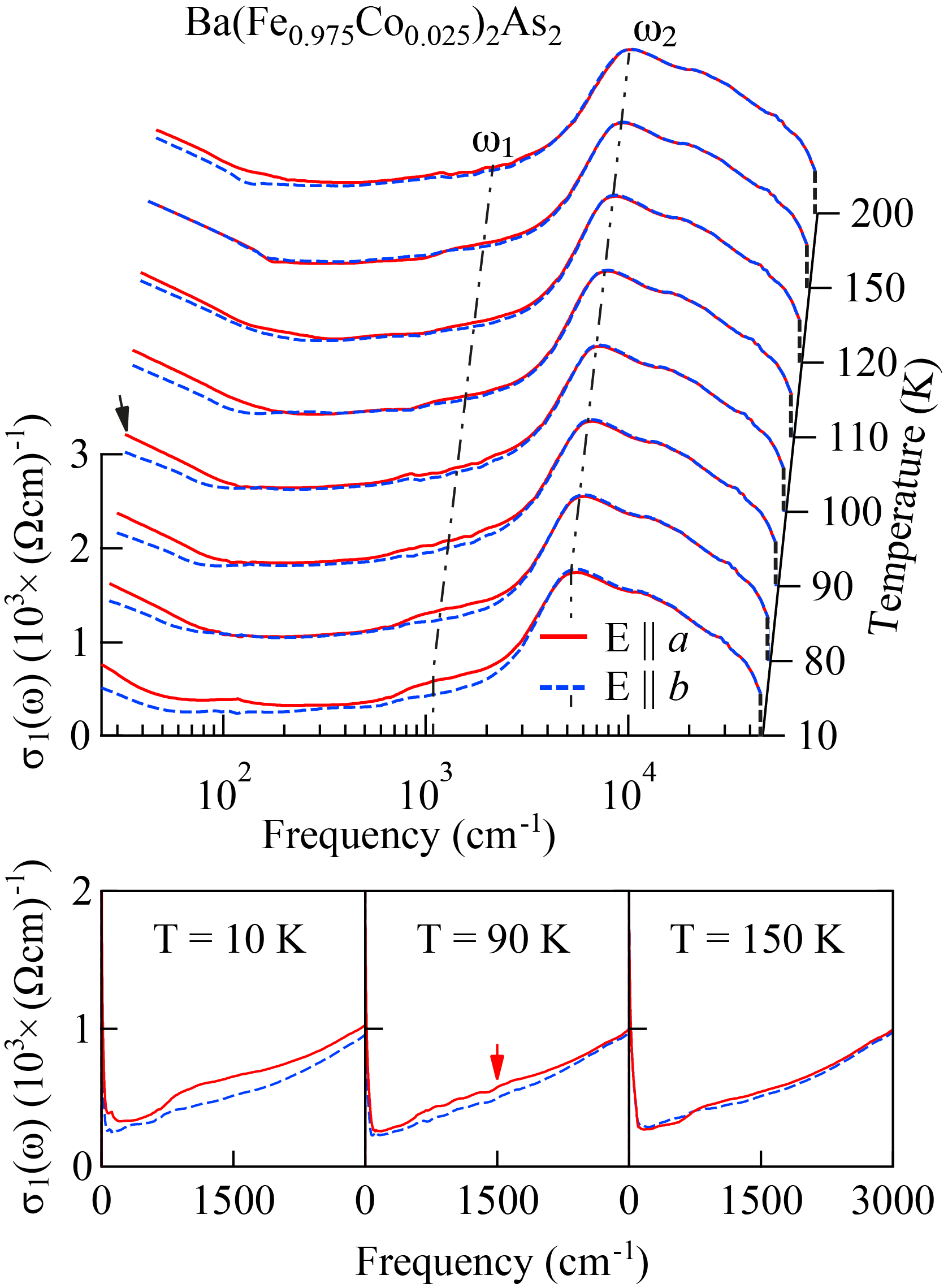}}
\subfigure[]{\includegraphics[width=5.91cm]{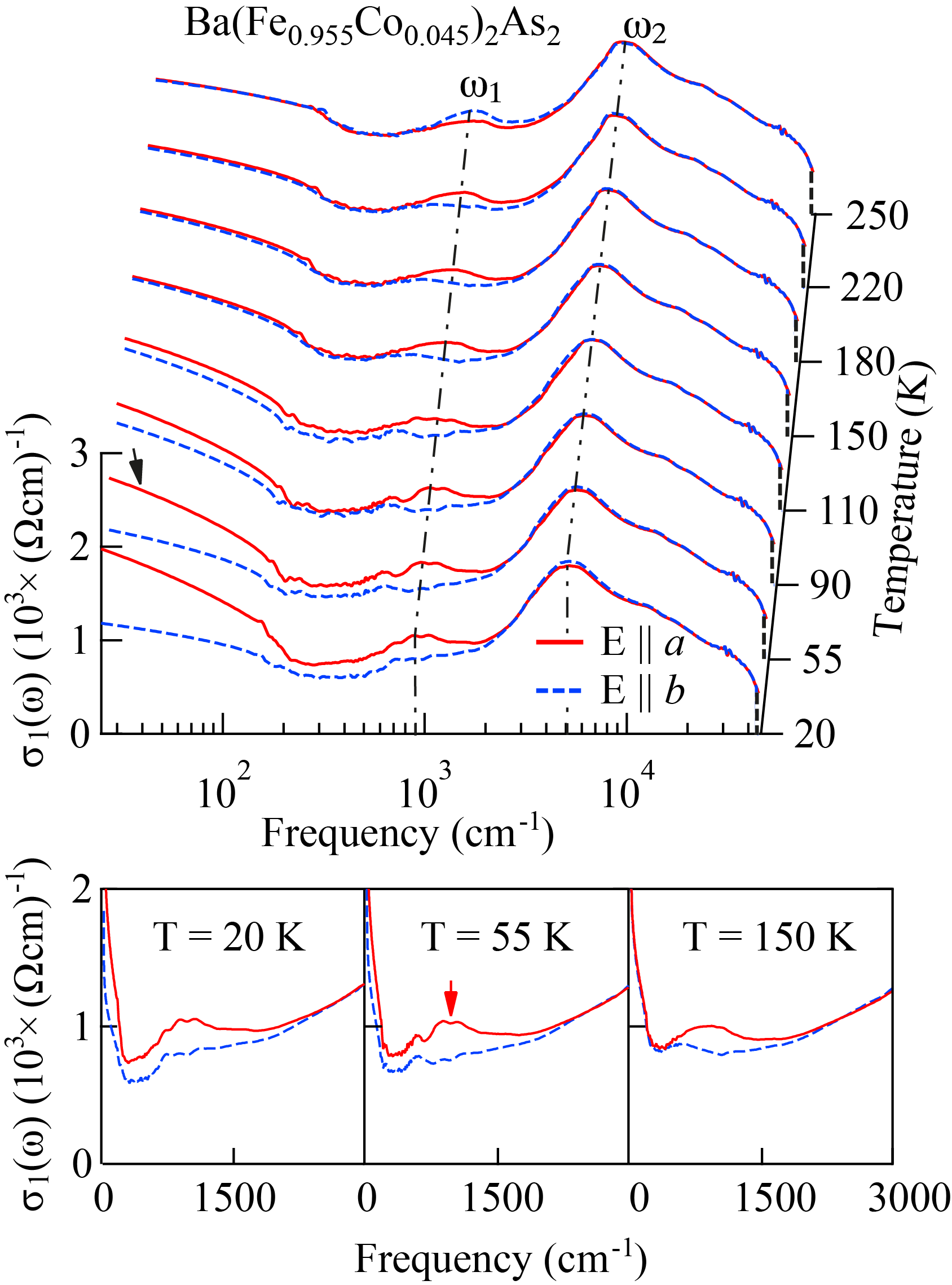}}
\caption{(color online) Temperature dependence of the optical conductivity of detwinned  BaFe$_2$As$_2$, Ba(Fe$_{0.975}$Co$_{0.025}$)$_2$As$_2$ and Ba(Fe$_{0.955}$Co$_{0.045}$)$_2$As$_2$ in the whole measured spectral range for two polarizations, parallel to the a-axis direction (E$\parallel$$a$, red solid line) or perpendicular to it (E$\parallel$$b$, blue dashed line). The black arrows indicate $\sigma_{1}(\omega)$ at the temperature close to the structural phase transition at $T_s$, while the red arrows (bottom panels) indicate the center of the MIR-band close to $T_N$. The dashed-dotted and dashed double-dotted line (top panels) mark the frequency $\omega_1$ and $\omega_2$ (see text).} \label{Sig1}
\end{figure*}
\section{Results}
%
%
%
%
%
\subsection{Reflectivity}
The three investigated compositions display overall similar features in their optical response but their polarization and temperature dependences show small but significant differences as we clarify in the presentation and discussion of our results below. Figure \ref{Ref} presents the optical reflectivity $R(\omega)$ in the whole measured frequency range of detwinned Ba(Fe$_{1-x}$Co$_x$)$_2$As$_2$ ($x = 0$, $x = 0.025$ and $x = 0.045$) at different temperatures and for the two polarization directions E$\parallel$$a$ and E$\parallel$$b$. As already recognized in the twinned (i.e., unstressed) specimens \cite{lucarelli}, $R(\omega)$ gently increases from the UV to the MIR region displaying an overdamped-like behavior. 
Below the MIR energy range, $R(\omega)$ gets progressively steeper with a sharp upturn at frequencies lower than 200 cm$^{-1}$ (Fig. \ref{Ref}). Close to zero frequency, $R(\omega)$ consistently merges in the HR extrapolations calculated with the $\sigma_{dc}$ values from Ref. \cite{chudw}. 
For all measured dopings we observed a polarization and temperature dependence of $R(\omega)$ from the FIR up to the MIR-NIR range, while between 5000 and 6000  cm$^{-1}$ the $R(\omega)$ spectra tend to merge together. The optical anisotropy is rather pronounced at low temperatures in the FIR region, when approaching the zero frequency limit. Interestingly for $x=0$, $R(\omega)$ increases with decreasing temperature along the $a$-axis, in agreement with the metallic character of the $dc$ transport properties \cite{chudw}. On the contrary, at low temperatures along the $b$-axis, there is first a depletion of $R(\omega)$ in the FIR energy range below 700 cm$^{-1}$ and then a steeper upturn below 100 cm$^{-1}$, consistently merging with the HR extrapolation. While somehow common to all compositions this depletion is less pronounced at higher doping levels such as $x = 0.025$ and $x = 0.045$, when entering the magnetic ordered phase at $T<T_{N}$ (Fig. \ref{Ref}b and \ref{Ref}c, respectively), similarly to what has been previously observed in the twinned specimens \cite{lucarelli}.

The anisotropy, discussed so far in the $dc$-limit of $R(\omega)$ at low temperatures, persists up to temperatures well above all phase transitions for the crystals held under uniaxial pressure, predominantly in the MIR-NIR region than in the FIR one. The black arrow in Fig. 3 highlight the $R(\omega)$ spectra collected at temperatures close to the respective $T_s$. The MIR-band centered at about 1500 cm$^{-1}$ (dotted-dashed lines and red arrow in Fig. 3) is of particular interest for both its temperature and polarization dependence. For $x=0$, we observe an interchange in the polarization dependence of $R(\omega)$ with decreasing temperature (Fig. \ref{Ref}a bottom panels). Such an interchange exactly occurs at the coupled magnetic and structural phase transition at 135 K. At higher doping levels ($x = 0.025$ and $x = 0.045$) the MIR-band observed at 1500 cm$^{-1}$ for $x=0$ is shifted towards lower frequencies  (red arrows in bottom panels of Fig. \ref{Ref}). Above 150 K the anisotropy of $R(\omega)$ in the MIR region is strongly reduced, but differently from the parent compound the $a$-axis spectrum of $x = 0.025$ and $x = 0.045$ is constantly above the $b$-axis one at all temperatures (Fig. \ref{Ref}b and \ref{Ref}c bottom panels).
%
%
%
%
\subsection{Optical Conductivity}
Figure \ref{Sig1} shows the real part $\sigma_{1}(\omega)$ of the optical conductivity of detwinned Ba(Fe$_{1-x}$Co$_x$)$_2$As$_2$ ($x = 0$, $x = 0.025$ and $x = 0.045$) for different measured temperatures along both polarization directions. In the visible and UV energy interval $\sigma_{1}(\omega)$  is characterized by polarization independent broad absorption bands which overlap with a dominant NIR contribution peaked at about 5000 cm$^{-1}$. Consistently with previous measurements on twinned samples \cite{lucarelli}, these components of $\sigma_{1}(\omega)$  are generally ascribed to the electronic interband transitions. These features in the UV-NIR range are not substantially altered by changing temperature, polarization, or doping level. 

As already observed for $R(\omega)$, the temperature and doping dependent optical anisotropy in $\sigma_{1}(\omega)$ is mainly evident in the FIR and MIR regions. In the FIR region, there is a strong polarization dependence of the itinerant charge carriers contribution to $\sigma_1(\omega)$. Along the $a$-axis $\sigma_{1}(\omega)$ shows a more pronounced metallic behavior which gets enhanced below $T_N$. Along the $b$-axis $\sigma_{1}(\omega)$ below $T_N$ is depleted due to the formation of a pseudogap, prior to displaying a metallic-like upturn for $\omega\rightarrow 0$. As formerly observed for the twinned samples \cite{lucarelli} this depletion of $\sigma_{1}(\omega)$ along the $b$-axis is less evident when increasing the doping.

The strong absorption peak dominating $\sigma_{1}(\omega)$ at about 5000 cm$^{-1}$ develops into a pronounced shoulder on its MIR frequency tail at about 1500 cm$^{-1}$. As anticipated in the presentation of the $R(\omega)$ data, this latter MIR-band in $\sigma_{1}(\omega)$ shows a strong polarization and doping dependence, as highlighted in Fig. 4 (bottom panels). One recognizes again the already mentioned interchange in the polarization dependence when crossing the structural transition for $x=0$ and its absence at higher dopings. Above $T_N$ in the MIR range, $\sigma_{1}(\omega)$ of $x=0.025$ and $x=0.045$ along the $a$-axis direction is above the $b$-axis values and remains constantly above it also below $T_N$ (top and bottom panels of Fig. \ref{Sig1}b and \ref{Sig1}c, respectively). Interestingly enough, for increasing doping the maximum of the MIR-band shifts to lower frequencies indicating that the MIR-band is significantly affected by the doping (dotted-dashed lines in top panels and red arrows in the bottom panels of Fig. \ref{Sig1}). 

The anisotropy in the optical response for the magnetic state can be anticipated by $ab-initio$ calculations based on density-functional-theory (DFT) as well as dynamical mean-field theory (DMFT) \cite{yin,sanna,sugimoto}. It was first shown that the optical anisotropy of the magnetic state, not present within the local spin density approximation, may result from DMFT-correlation \cite{yin}. Alternatively, DFT-calculations of the optical conductivity within the full-potential linear augmented plane-wave (LAPW) method reproduce most of the observed experimental features, in particular an anisotropic magnetic peak located at about 0.2 eV (1600 cm$^{-1}$), which was ascribed to antiferromagnetically ordered stripes \cite{sanna}. The optical anisotropy, as observed experimentally, was even shown to agree with the solution of a three dimensional five-orbital Hubbard model using the mean-field approximation in the presence of both orbital and magnetic order \cite{lv_anis}. Moreover, it has been recently pointed out that interband transitions, whose relevance is manifested by first-principle calculations, give a non negligible contribution already in the infrared region, spanning the experimental energy interval of the MIR-band \cite{benfatto}. We will return later on to the comparison between experiment and theory.
%
%
%
%
\subsection{Fits}
\begin{figure}[!tb]
\center
\includegraphics[width=8cm]{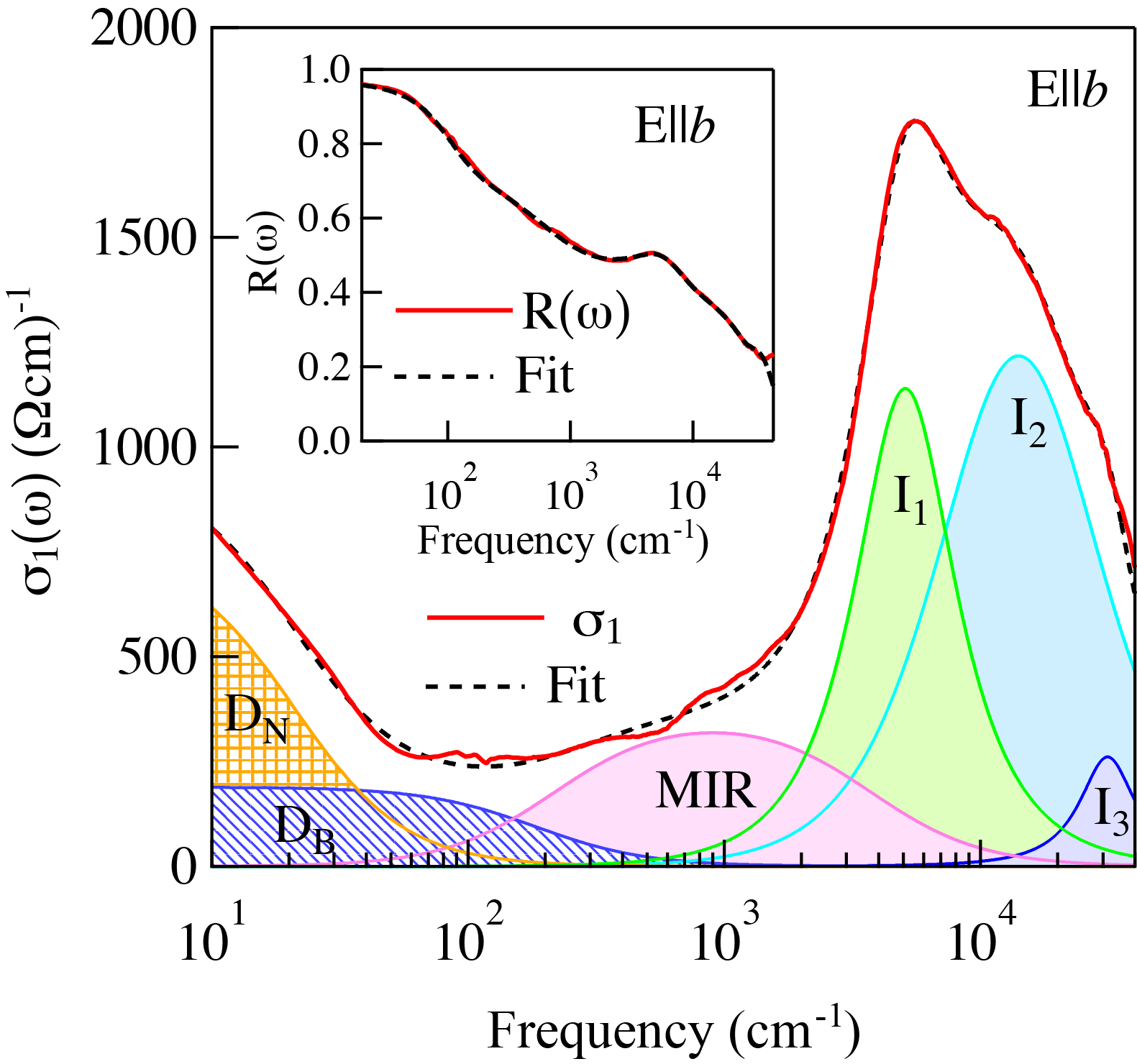}
\caption{(color online) Optical conductivity along the $b$-axis at 10 K of detwinned Ba(Fe$_{0.975}$Co$_{0.025}$)$_2$As$_2$ compared with the total Drude-Lorentz fit (black dashed line) and the corresponding components: the narrow (D$_N$) and broad (D$_B$) Drude terms, the mid-infrared (MIR) band and the oscillators ($I_1$, $I_2$, $I_3$) fitting the interband transitions. The inset shows the comparison between the measured reflectivity $R(\omega)$ and the resulting fit.} \label{Fits}
\end{figure}
\begin{figure*}[ht]
\centering
\subfigure[]{\includegraphics[width=5.91cm]{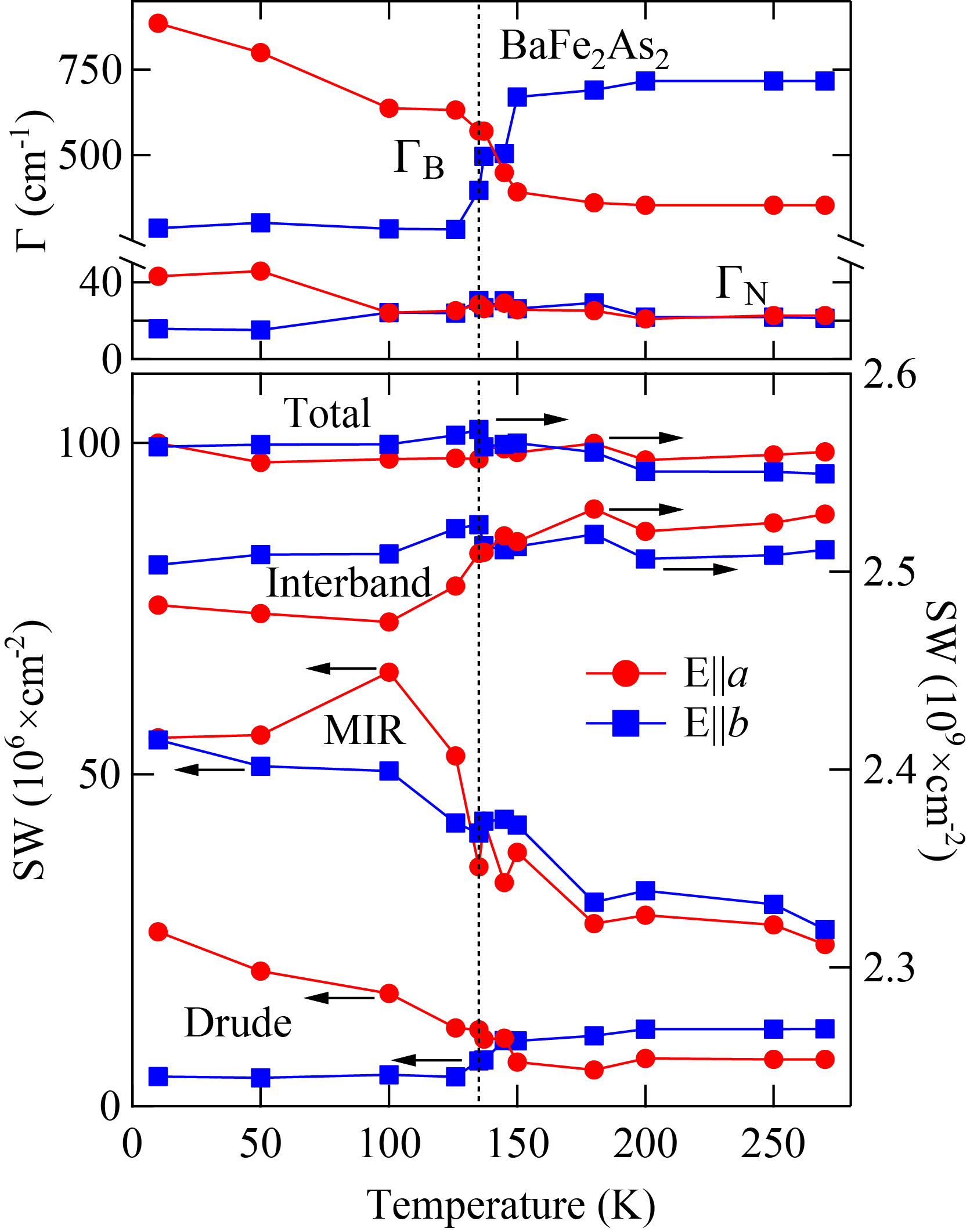}}
\subfigure[]{\includegraphics[width=5.91cm]{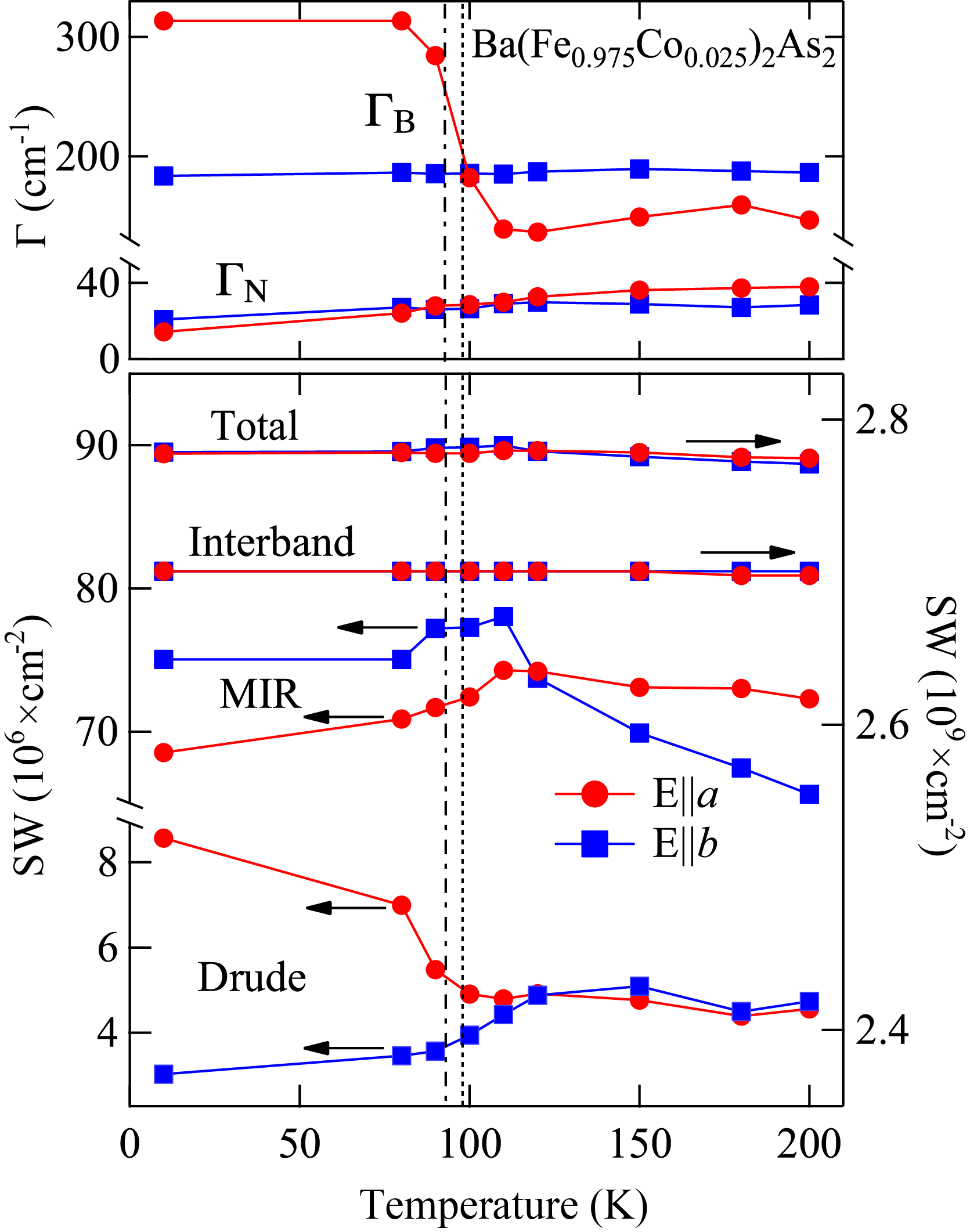}}
\subfigure[]{\includegraphics[width=5.91cm]{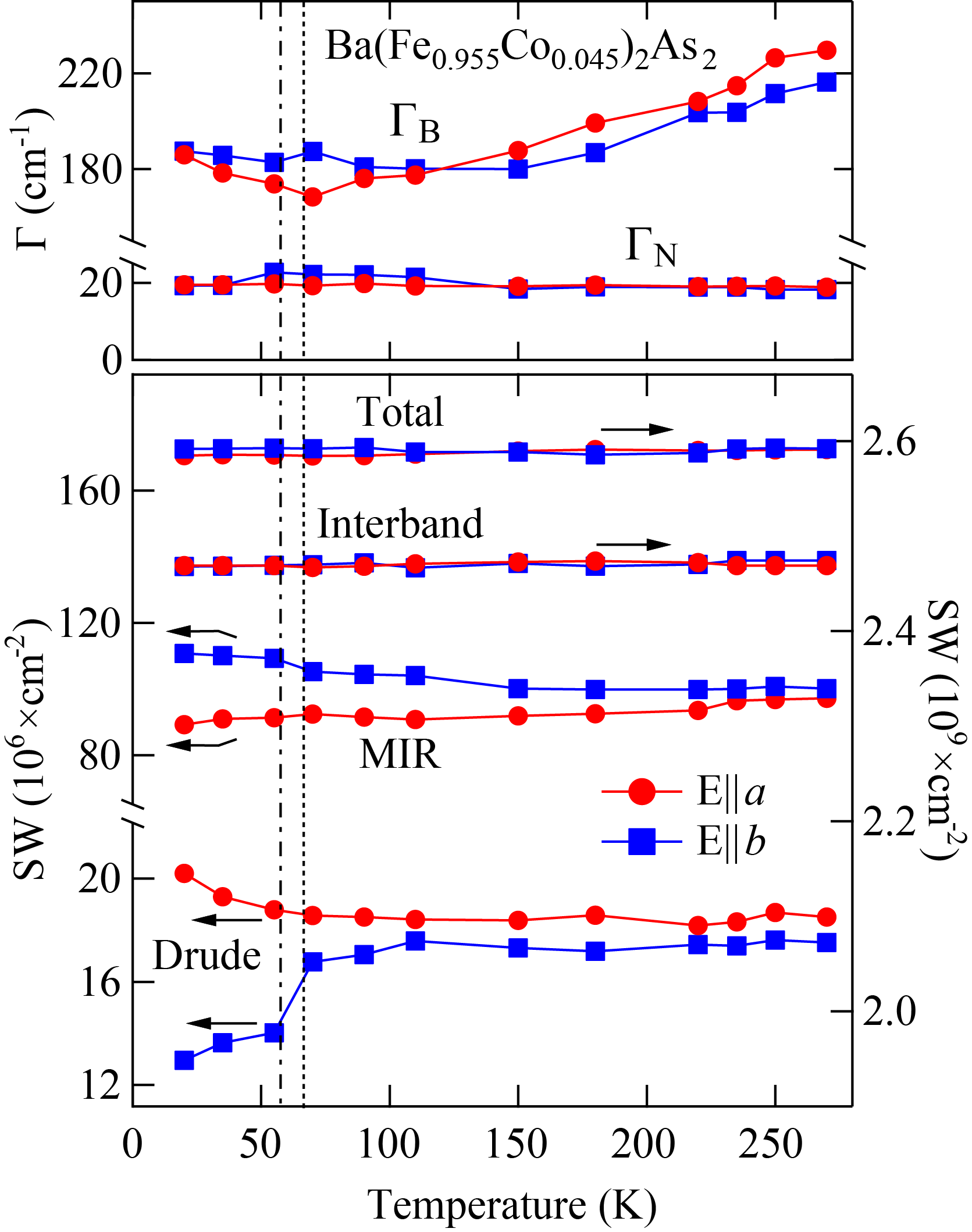}}
\caption{(color online) Temperature and polarization dependence of the widths for the broad ($\Gamma$$_{B}$) and narrow ($\Gamma$$_{N}$) Drude terms (top panels) and the spectral weight (SW) (bottom panels) for the Drude, the MIR and the interband components compared with the total SW obtained from the fits of BaFe$_2$As$_2$, Ba(Fe$_{0.975}$Co$_{0.025}$)$_2$As$_2$ and Ba(Fe$_{0.955}$Co$_{0.045}$)$_2$As$_2$. The dotted and dotted-dashed lines indicate the structural and magnetic transitions at $T_s$ and $T_N$, respectively.} \label{SW_Gamma}
\end{figure*}
In order to study the various contributions shaping the optical conductivity at different energies, we apply the well-established phenomenological Drude$-$Lorentz approach. Consistent with our previous investigations on twinned samples \cite{lucarelli}, we ascribe two Drude contributions (one narrow and one broad) to the effective metallic part of $\sigma_{1}(\omega)$ and a series of Lorentz harmonic oscillators (h.o.) for all excitations (phononic and electronic) at finite frequencies. Figure \ref{Fits} presents all the fitting components for $x = 0.025$ along the $b$-axis direction measured at 10 K, acting here as a representative example. The use of two Drude components in the fit procedure phenomenologically mimics the multi-band scenario and implies the existence of two electronic subsystems as revealed for a wide range of iron-pnictide compounds \cite{wu}. The narrow Drude term is relevant at very low frequencies and it is obviously tied to the necessary HR extrapolation of $R(\omega)$ for $\omega\rightarrow$0. The broad one acts as a background of $\sigma_1(\omega)$ and dominates the optical conductivity up to the MIR energy interval. As we shall elaborate later on, both Drude terms contribute to the total $dc$ conductivity. Besides the Drude terms, we chose one broad h.o. for the temperature dependent MIR-band and three broad h.o.'s ($I_1$, $I_2$ and $I_3$ in Fig. \ref{Fits}) for the strong absorption featuring the broad peak centered at about 5000 cm$^{-1}$. The complex dielectric function $\tilde{\varepsilon}=\varepsilon_1(\omega)+i\varepsilon_2(\omega)$ can be expressed as follows:
\begin{align}\nonumber
\tilde{\varepsilon}=\varepsilon_{\infty}-\frac{\omega^2_{PN}}{\omega^2-i\omega\Gamma_N}-\frac{\omega^2_{PB}}{\omega^2-i\omega\Gamma_B}+ \\
+\frac{S^2_{MIR}}{\omega^2_{MIR}-\omega^2-i\omega\gamma_{MIR}}+
\sum_{j=1}^3\frac{S^2_j}{\omega^2_j-\omega^2-i\omega\gamma_j}
\end{align}
where $\varepsilon_{\infty}$ is the optical dielectric constant, $\omega^2_{PN}$, $\omega^2_{PB}$ and $\Gamma$$_{N}$, $\Gamma$$_{B}$ are respectively the plasma frequencies, defined as $\omega^2_{P}=\frac{4\pi e^2n}{m^*}$, and the widths of the narrow and broad Drude peaks. The latter parameters represent the scattering rates of the itinerant charge carriers, of which $n$, $m^*$ and $e$ are then the density, the effective mass and the charge, respectively. The parameters of the $j$th Lorenz h.o. as well as those of the MIR-band are: the center-peak frequency ($\omega_{j}$ and $\omega_{MIR}$), the width ($\gamma_{j}$ and $\gamma_{MIR}$) and the mode strength ($S_j^2$ and $S_{MIR}^2$). The fit constraints are such that the measured reflectivity and the real part of the optical conductivity are simultaneously reproduced by the identical set of fit-parameters \cite{lucarelli}, which reduces the degree of freedom in the parameters choice. The upper boundary for the temperature dependence of the optical conductivity is found to be close to the NIR peak in $\sigma_1(\omega)$ at about 5000 cm$^{-1}$. Thus, for all temperatures and dopings we fit $R(\omega)$ and $\sigma_1(\omega)$ by varying the parameters for both Drude terms, the MIR-band and the $I_1$ h.o., while keeping constant the parameters associated to the two high frequency oscillators $I_2$ and $I_3$. We systematically adopted our fitting procedure (Fig. 5) for both polarization directions and for all measured temperatures from 10 K to 270 K of the studied compositions. The only exception, however, is for the MIR-band of $x=0$ along the $b$-axis, where we added one more component, in order to achieve a better fit quality. We checked that this fit-variation has negligible impact on the overall trend of the extracted parameters. The remarkable agreement of the fitting results with the measured reflectivity and optical conductivity (see e.g. Fig. \ref{Fits}) further demonstrates the overall good quality of the fits. Therefore, we are confident that the fit results allow us to identify robust trends in relevant physical parameters. 
\section{Discussion}
%
%
%
%
%
We can first exploit our phenomenological fits from the perspective of a so-called spectral-weight ($SW$) analysis, represented by the squared plasma frequencies and mode strengths of the Drude terms and Lorentz h.o.'s, respectively. The Drude-Lorentz procedure thus allows us to disentangle the redistribution of $SW$ in selected energy intervals and tells us how the same $SW$ is reshuffled among the various components as a function of temperature. Of particular interest is the total Drude weight given by $SW_{Drude}=\omega_{PN}^2+\omega_{PB}^2$. Furthermore, we consider the $SW$ encountered in the MIR-band and electron interband transitions, defined as $SW_{MIR}=S_{MIR}^2$ and $SW_{Int}=\sum_{j=1}^{3}S_{j}^2$, respectively. The total spectral weight is then given by the area under the conductivity spectrum and  can be expressed as \cite{grunerbook}: $SW_{Total}=SW_{Drude}+SW_{MIR}+SW_{Int}$. Due to the high fit-quality of $\sigma_1(\omega)$ for the chosen components, $SW_{Total}$ is also equivalent to $\int^{\omega_c}_0\sigma_1(\omega)d\omega$, where $\omega_c$ corresponds to a cutoff frequency and basically sets the upper frequency limit of our measurements. This implies that, if $SW_{Total}$ does not change with temperature within the energy interval between zero and $\omega_c$, any redistribution of the spectral weight will fully occur among the fitting components. 
\begin{figure*}[ht]
\center
\subfigure[]{\includegraphics[width=5.91cm]{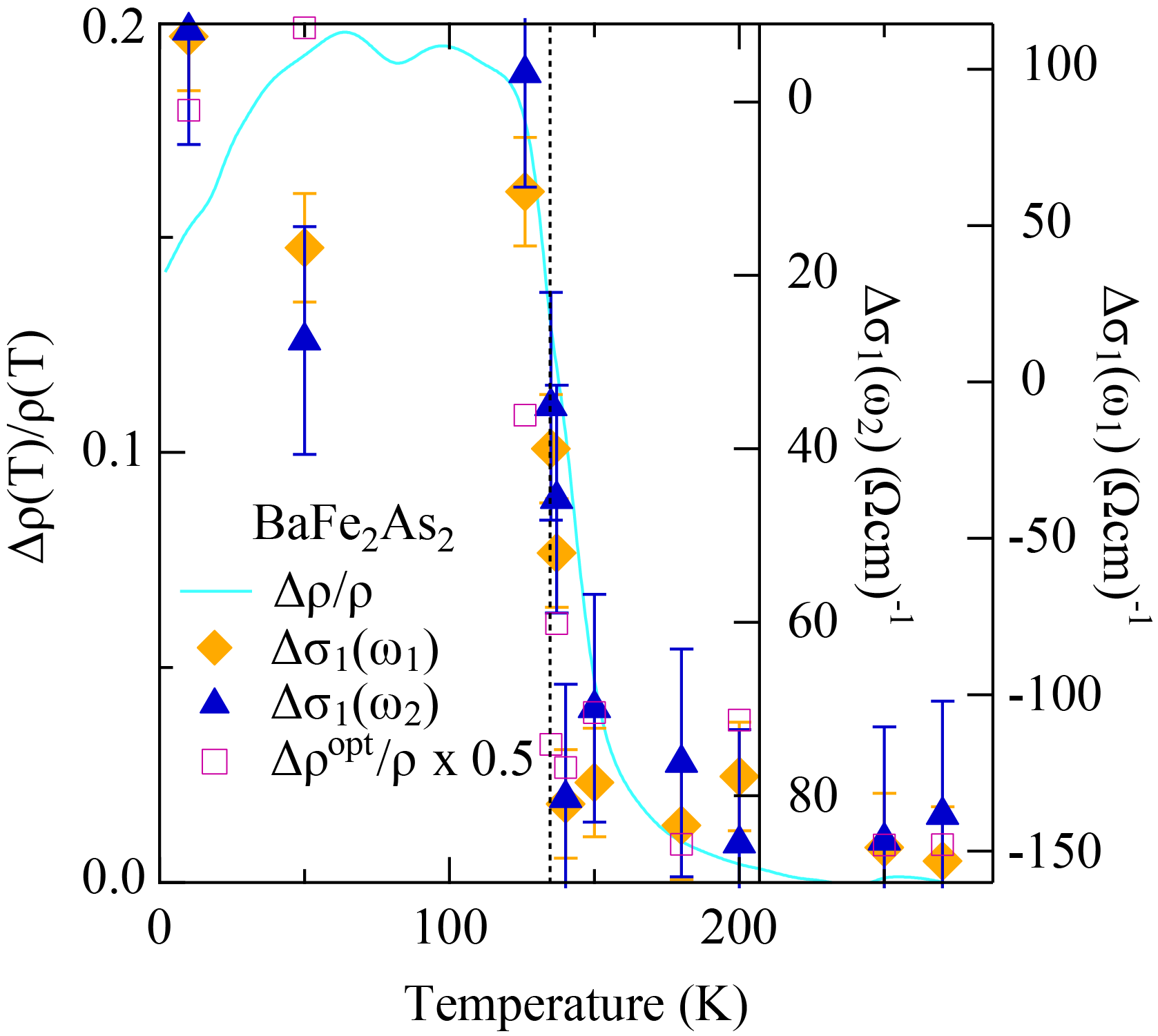}}
\subfigure[]{\includegraphics[width=5.91cm]{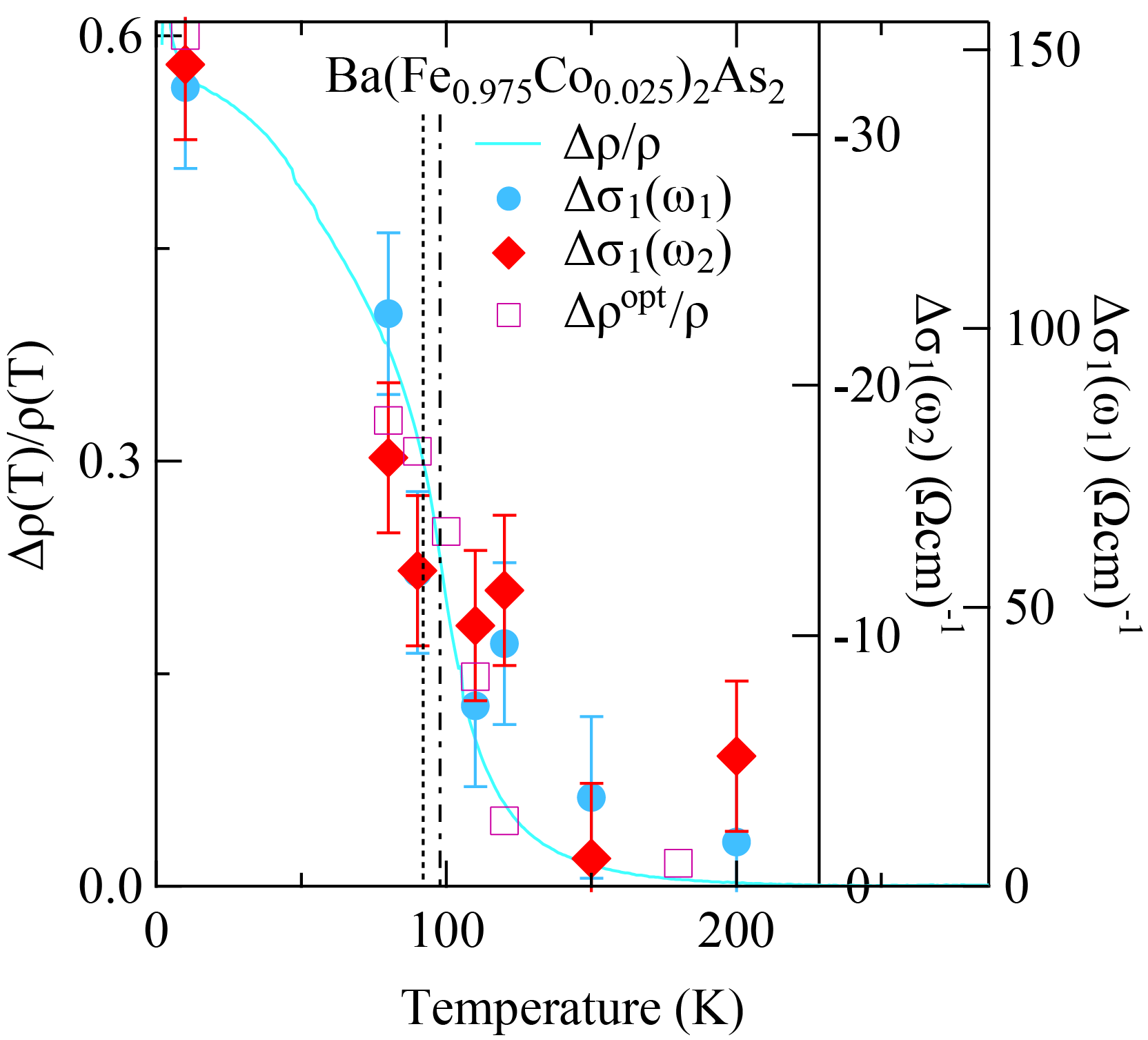}}
\subfigure[]{\includegraphics[width=5.91cm]{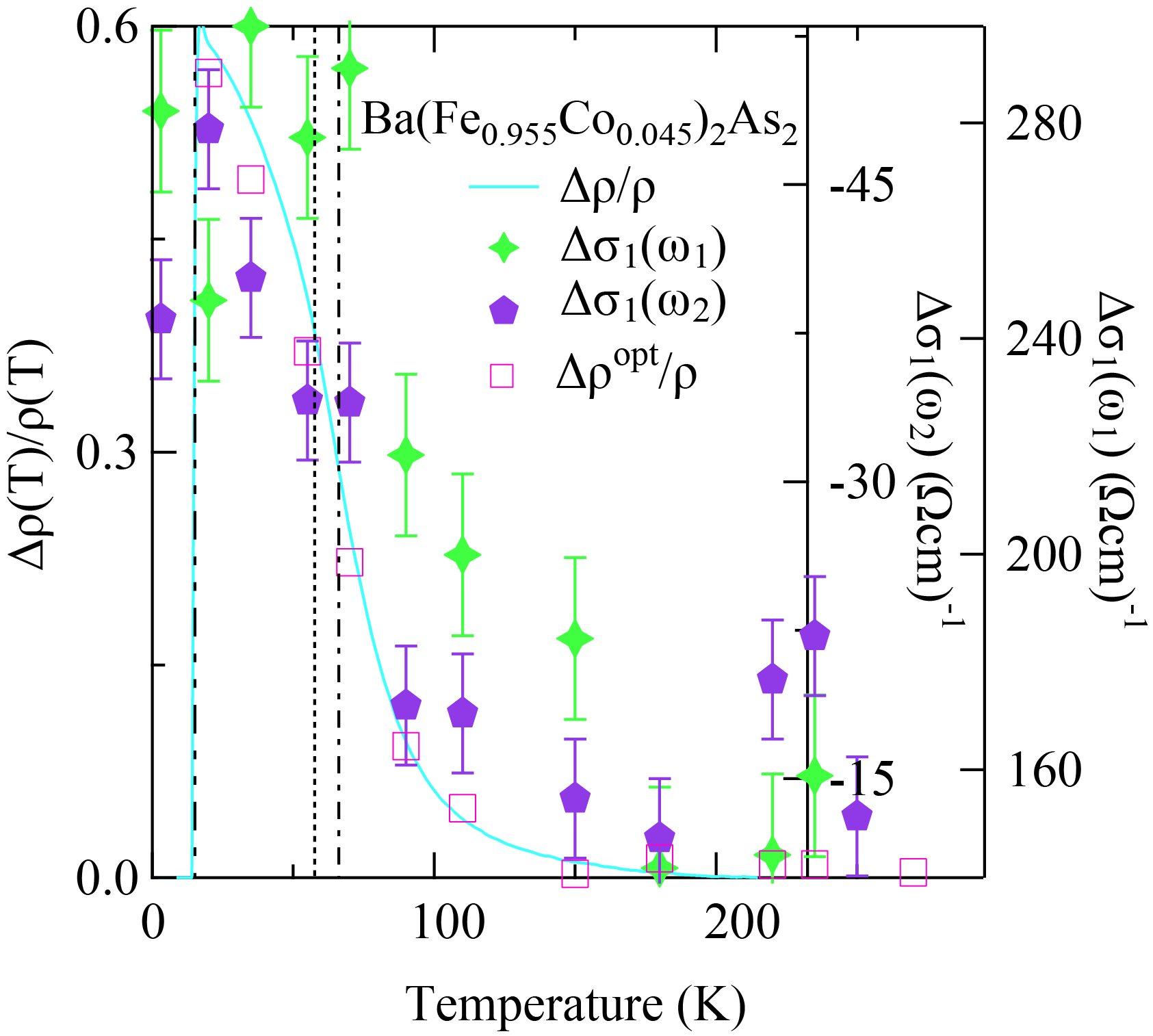}}
\caption{(color online) Temperature dependence of the dichroism $\Delta$$\sigma$$_1$($\omega$) of BaFe$_2$As$_2$, Ba(Fe$_{0.975}$Co$_{0.025}$)$_2$As$_2$ and Ba(Fe$_{0.955}$Co$_{0.045}$)$_2$As$_2$ at $\omega$$_1$ and $\omega$$_2$ (Fig. \ref{Sig1}) compared to $\frac{\Delta\rho}{\rho}$ obtained from the $dc$ transport data, as well as from the Drude terms in $\sigma$$_1$($\omega$)  ($\Delta$$\rho$$^{opt}$/$\rho$ ). The vertical dotted-dashed, dotted and double-dotted-dashed lines mark the structural, magnetic and superconducting phase transitions at $T_s$, $T_N$ and $T_c$, respectively.} \label{Anis_Ratio}
\end{figure*}

The lower panels of Fig. \ref{SW_Gamma} show the temperature dependence of $SW_{Total}$ and the spectral weight associated to the two Drude terms, the MIR-band and the three interband excitations ($I_i$). $SW_{Total}$ stays constant with temperature along both polarization directions for all doping values thus satisfying the $f$-sum rule \cite{grunerbook}. When crossing the structural and magnetic transitions, the spectral weight of the parent compound along the $a$-axis is redistributed from high to low frequencies, since the interband transition components loose weight in favor of the Drude and MIR-band (bottom panel of Fig. \ref{SW_Gamma}a). On the contrary along the $b$-axis, the Drude term looses $SW$ for $T<T_N$ in favor of the MIR-band. At higher dopings ($x = 0.025$ and 0.045) the overall $SW_{Int}$ shows less pronounced temperature dependence for both lattice directions. The spectral weight variations with temperature mainly occur between the Drude terms and the MIR-band for both directions. The resulting temperature dependence of $SW_{Drude}$, increasing along the $a$- but decreasing along the $b$-axis across $T_N$ and $T_s$ for all dopings, is rather intriguing and unanticipated, the significance of which with respect to the $dc$ transport properties will be addressed below. For the time being, it is worth emphasizing that the Drude weight anisotropy may strongly depend on the topology and morphology of the reconstructed Fermi surface below $T_s$ (i.e., anisotropy of the Fermi velocity), as evinced from a five-band Hamiltonian at the mean-field level \cite{valenzuela}. Of interest is also for $x=0$ the temperature independent total Drude weight, which is larger along the $b$-axis than along the $a$-axis for $T>T_s$, thus inverting the polarization dependence observed below $T_s$. This astonishing behavior for $x=0$ might be compatible with a recent multi-orbital model \cite{lv_anis}. Given the ARPES results \cite{zxshen}, showing that the band splitting diminishes with increasing temperature, one would have eventually anticipated that the Drude weight gets indeed isotropic above $T_s$. Nonetheless, the Drude weight anisotropy above $T_s$ is suppressed upon doping (Fig. \ref{SW_Gamma}). Experimentally, such a trend of $SW_{Drude}$ remains to be verified under controlled uniaxial pressure conditions, while theoretically it awaits confirmation within doping-dependent models. 

From the width at half maximum of the Drude resonance we can extract the scattering rates of the itinerant charge carriers. The broad ($\Gamma_{B}$) and narrow ($\Gamma_{N}$) Drude widths for $x = 0$, 0.025 and 0.045 are shown in the top panels of Fig. \ref{SW_Gamma}. An anisotropy in the scattering rate is most evident for $x = 0$ and $x = 0.025$. For $x = 0$ the Drude scattering rates $\Gamma_{N}$ and $\Gamma_{B}$ increase along the $a$-axis and decrease along the $b$-axis for decreasing temperatures across the phase transitions. Above $T_s$ for $x=0$, $\Gamma_N$ along both axes saturates to the identical constant value, while $\Gamma_B$ displays an inversion in the polarization dependence with respect to the situation below $T_s$ (Fig. 6, upper panel), before saturating to temperature independent values. For $x=0.025$ only $\Gamma_B$ along the $a$-axis undergoes a sudden incremental change at $T_N$, while all other scattering rates remain almost constant. $\Gamma_B$ above $T_s$ is switched with respect to below $T_s$ between the two polarization directions, similarly to $x=0$ yet less pronounced. For $x = 0.045$ the $\Gamma_B$ scattering rates display a more metallic behavior with decreasing temperature above $T_N$ and an almost negligible polarization dependence. Below $T_N$ there is a weak upturn to higher values, particularly along the $a$-axis. $\Gamma_N$ remain constant at all temperatures and for both directions.

The overall temperature dependence of the scattering rates below $T_N$ as evinced from the analysis of the optical response is expected with respect to the well-established magnetic order \cite{li}. Particularly for $x=0$ and $x=0.025$, the larger scattering rates along the elongated antiferromagnetic $a$-axis than along the shorter ferromagnetic $b$-axis for temperatures below $T_N$ may arise because of reduced hopping or of scattering from spin-fluctuations with large momentum transfer (i.e., by incoherent spin waves) \cite{turner,devereaux}. The anisotropic scattering rate in the paramagnetic state, at least for $x=0$ and $x=0.025$, might also be in agreement with predictions based on interference between scattering by impurities and by critical spin fluctuations in the Ising nematic state \cite{fernandes2}. Similarly to our previous discussion on the Drude weight, we shall caution the readership, that the trend in the scattering rates, particularly above $T_s$, as well as their doping dependence should be also verified with tunable uniaxial pressures, in order to guarantee equal experimental conditions. A comprehensive theoretical framework, approaching different temperature regimes and considering the impact of doping, is also desired.

Having determined the two parameters governing the $dc$ transport properties, it is worth pursuing at this point the compelling comparison between the temperature dependence of the optical anisotropy and the anisotropy ratio of the $dc$ transport properties, defined as $\frac{\Delta\rho}{\rho}$=$\frac{2(\rho_b-\rho_a)}{(\rho_b+\rho_a)}$ (Fig. \ref{Anis_Ratio}) \cite{devereaux}. From the Drude terms, fitting the effective metallic contribution of $\sigma_1(\omega)$ over a finite energy interval, we can estimate the $dc$ limit of the conductivity ($\sigma_0^{opt}=(\omega_p^N)^2/4\pi\Gamma_N+(\omega_p^B)^2/4\pi\Gamma_B$) more precisely than simply extrapolating $\sigma_1(\omega)$ to zero frequency. The anisotropy ratio $\frac{\Delta\rho^{opt}}{\rho}$, reconstructed from the optical data, is thus compared in Fig. \ref{Anis_Ratio} to the equivalent quantity from the transport investigation. The agreement in terms of $\frac{\Delta\rho}{\rho}$ between the optical and $dc$ investigation is outstanding for $x$=0.025 and 0.045 at all temperatures. For $x$= 0, $\frac{\Delta\rho^{opt}}{\rho}$ is nonetheless slightly larger than the $dc$ transport anisotropy for $T<T_s$. This disagreement might originate from a difference in the applied stress in the optical and $dc$ transport measurements, or from differences in scattering rate of samples used for the two types of measurements. 

Significantly, analysis of the optical properties for all compositions seems to indicate that anisotropy in the Fermi surface parameters, such as the enhancement(depletion) of the total Drude spectral weight occurring along the $a(b)$-axis, outweighs the (large at some compositions) anisotropy in the scattering rates (Fig. \ref{SW_Gamma}) that develops below $T_N$ in terms of the effect on the $dc$ transport properties (Fig. \ref{Anis_Ratio}). This is an important result from the optical investigation, which indeed enables to extract both pieces of information governing the behavior of the $dc$ transport properties.

%
%
%
%
%
\begin{figure*}[ht]
\centering
\subfigure[]{\includegraphics[height=6.12cm]{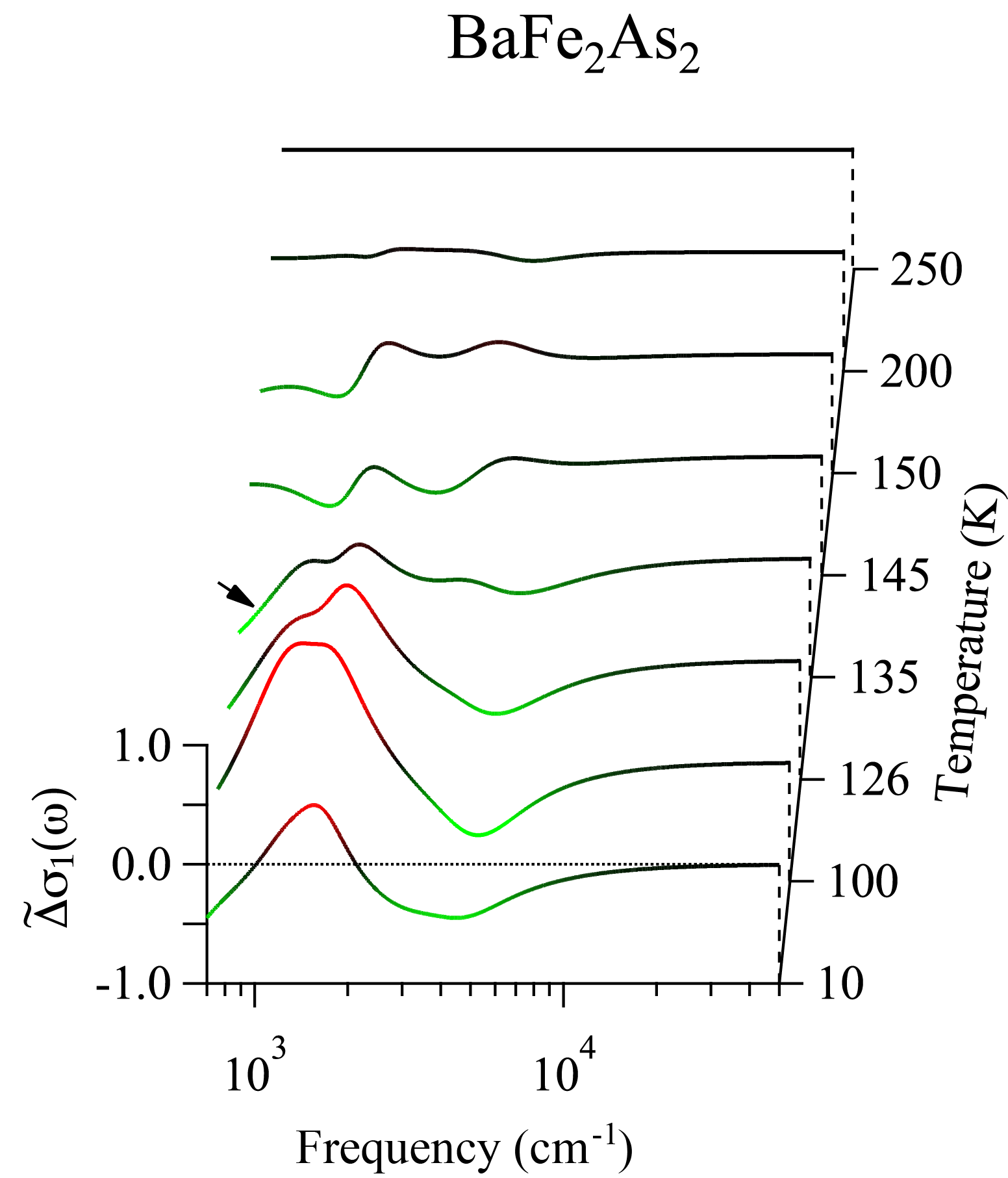}}
\subfigure[]{\includegraphics[height=6.12cm]{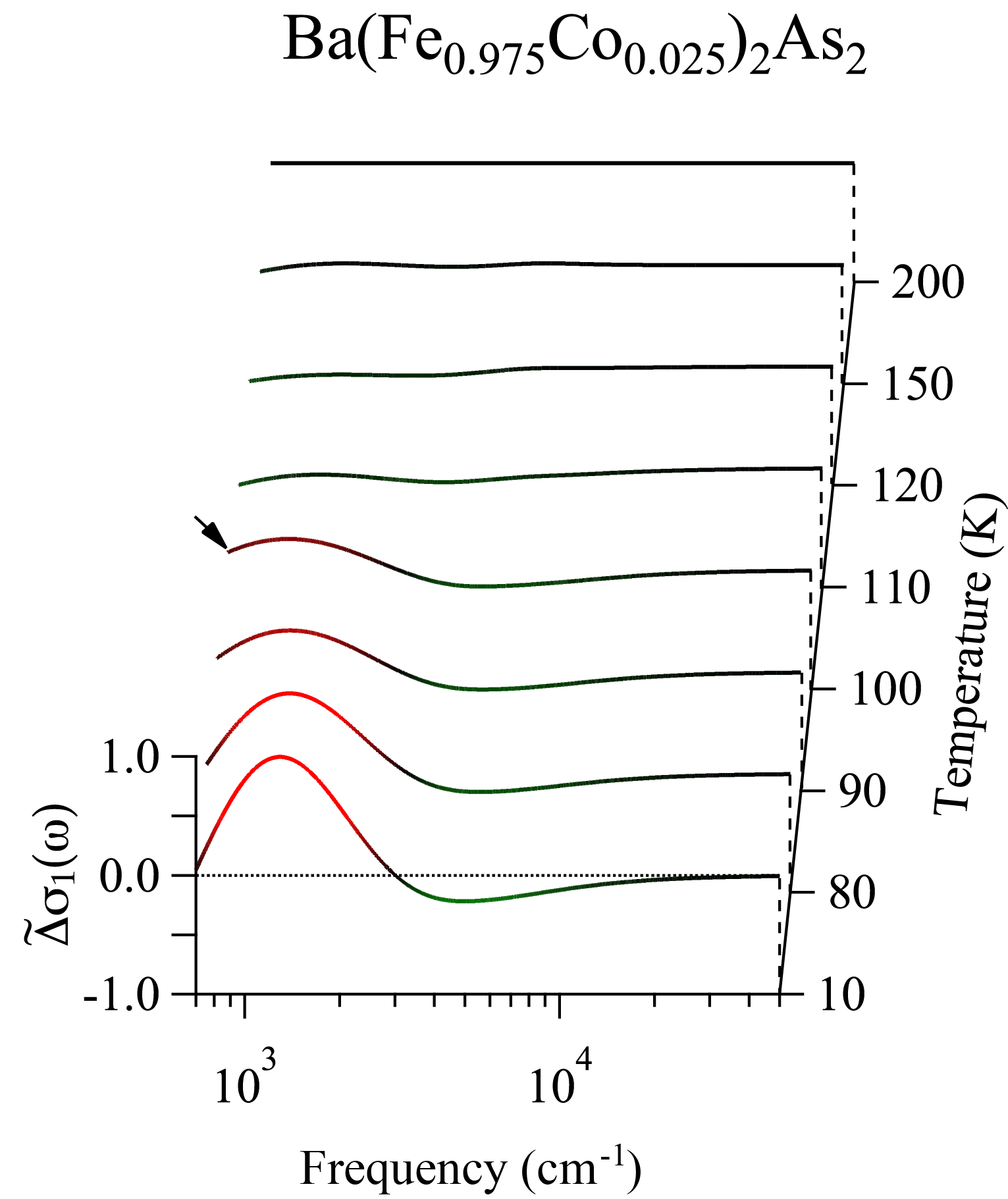}}
\subfigure[]{\includegraphics[height=6.12cm]{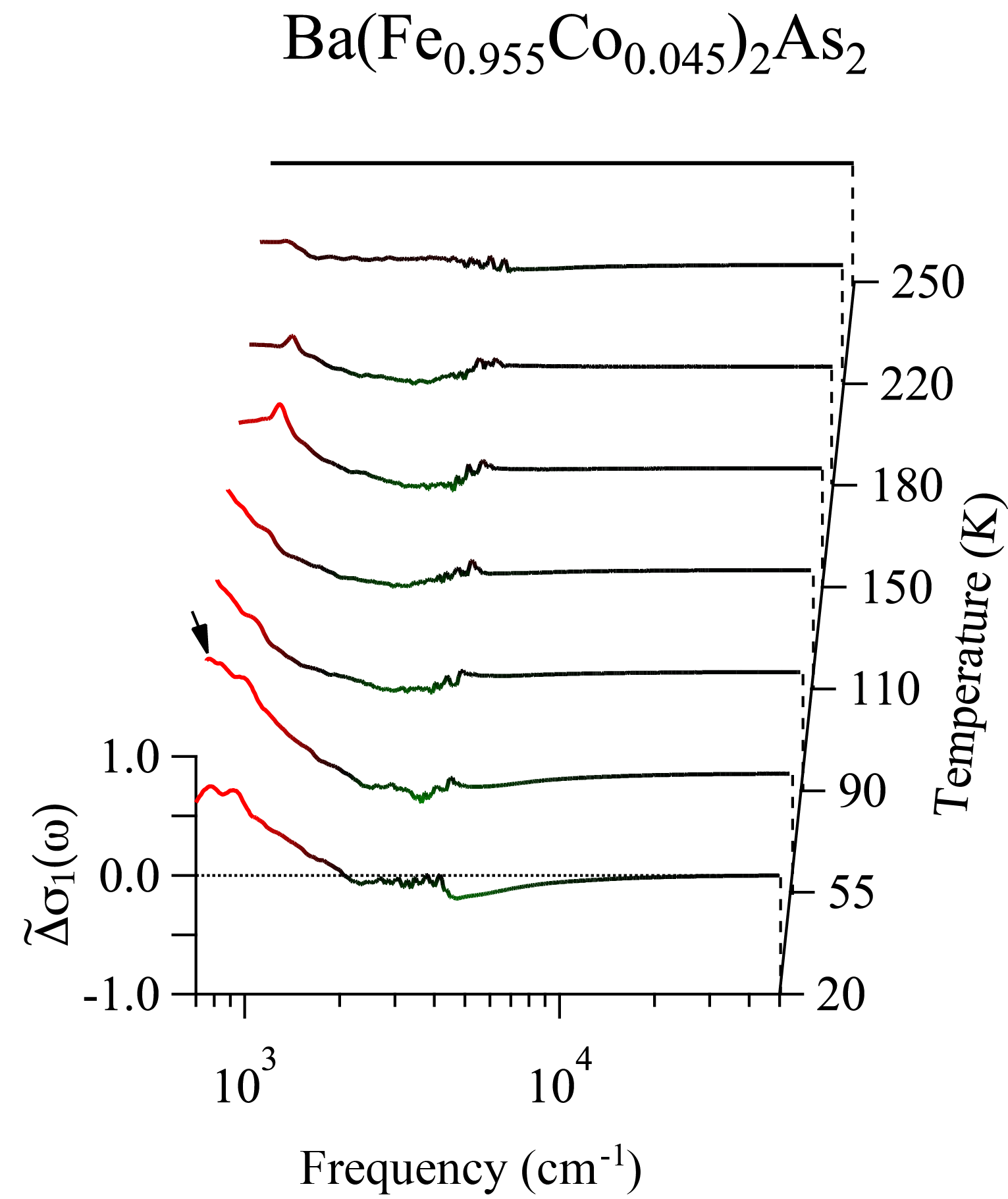}}
\caption{(color online) Temperature dependence of the normalized relative difference $\widetilde{\Delta}\sigma_1(\omega,T)=\Delta\sigma_1(\omega,T)-\Delta\sigma_1(\omega, 300 K)$ of the optical dichroism of BaFe$_2$As$_2$, Ba(Fe$_{0.975}$Co$_{0.025}$)$_2$As$_2$ and Ba(Fe$_{0.955}$Co$_{0.045}$)$_2$As$_2$. The black arrows indicate the $\Delta$$\sigma$$_1$($\omega$) at the temperature close to the structural phase transition at $T_s$. As color code: red marks positive, green negative values of $\widetilde{\Delta}\sigma_1(\omega,T)$.} \label{DeltaSig1}
\end{figure*}

In order to emphasize the relevant polarization dependence at high frequencies, we calculate the linear dichroism $\Delta\sigma_1(\omega)$, as defined in the Introduction. For the purpose of further enhancing the optical anisotropy, we show in Fig. \ref{DeltaSig1} $\widetilde{\Delta}\sigma_1(\omega,T)$, which is defined as $\Delta\sigma_1(\omega,T)$ from the MIR to the UV for $x = 0$, 0.025 and 0.045 at various temperatures after having subtracted its corresponding room temperature values and being appropriately normalized. The dichroism persists above $T_N$ in the MIR range (Fig. \ref{DeltaSig1}), pairing our direct observations in terms of $R(\omega)$ and $\sigma_1(\omega)$ (Fig. \ref{Ref} and \ref{Sig1}). This representation highlights once more that the MIR-feature moves towards lower frequencies upon increasing doping. 

It is especially interesting to compare the temperature dependence of the $dc$ ($\frac{\Delta\rho}{\rho}$) \cite{chudw} and optical ($\Delta\sigma_1(\omega)$) anisotropy \cite{dusza}. Two characteristic frequencies, identifying the position of the peaks in $\sigma_1(\omega)$ (Fig. \ref{Sig1}), are selected to follow the temperature dependence of $\Delta\sigma_1(\omega)$; namely, $\omega_1$=1500 cm$^{-1}$ and $\omega_2$=4300 cm$^{-1}$ for $x=0$; $\omega_1$=1320 cm$^{-1}$ and $\omega_2$=5740 cm$^{-1}$ for $x=0.025$; $\omega_1$=912 cm$^{-1}$ and $\omega_2$=5182 cm$^{-1}$ for $x=0.045$. It is remarkable that the temperature dependence of $\Delta\sigma_1(\omega)$ at $\omega_1$ and $\omega_2$ follows the temperature dependence of $\frac{\Delta\rho}{\rho}$ in both compounds (Fig. \ref{Anis_Ratio}). $\Delta\sigma_1(\omega_i)$ ($i=1,2$) saturates at constant values well above $T_s$ and then displays a variation for $T < 2T_s$. Here we first underscore that the rather pronounced optical anisotropy, extending up to temperatures higher than $T_s$ for the stressed crystals, clearly implies an important pressure-induced anisotropy in the electronic structure, which is also revealed by ARPES measurements \cite{zxshen,wang}. 

Since the dichroism directly relates to a reshuffling of spectral weight in $\sigma_1(\omega)$ in the MIR-NIR range (Fig. 4a and 4b), $\Delta\sigma_1(\omega)$ at $\omega_1$ is interrelated to that at $\omega_2$ (i.e., the right $y$-axis for $\Delta\sigma_1(\omega_i)$ in Fig. \ref{Anis_Ratio} are inverted between $\omega_1$ and $\omega_2$), so that the behavior of $\Delta\sigma_1(\omega)$ is monotonic as a function of temperature and opposite in sign between $\omega_1$ and $\omega_2$ (Fig. \ref{Anis_Ratio} and \ref{DeltaSig1}). For $x$=0.025 (Fig. \ref{Anis_Ratio}) $\Delta\sigma_1(\omega_i)$=0 at $T >> T_s$. However for $x$= 0 and 0.045,  $\Delta\sigma_1(\omega_i)$ is found to be constant but apparently different from zero for $T >> T_s$. The origin of the finite (but constant) dichroism at high temperatures for these samples is at present unclear, and might reflect a systematic effect due to imperfect experimental conditions (e.g., too strong applied uniaxial pressure). Nevertheless, the overall temperature dependence seems to behave in a very similar manner for all compositions. Significantly, the absolute variation of the dichroism across the transitions at selected frequencies is larger for $x$=0 than for $x$=0.025 and 0.045, contrary to the anisotropy in the $dc$ resistivity \cite{chudw}. This doping-dependence needs to be studied in a controlled pressure regime in order to exclude effects arising from different degrees of detwinning ($T<T_s$) and different magnitude of induced anisotropy ($T>T_s$). Even so, it is encouraging that, contrary to the $dc$ resistivity, the changes in the electronic structure appear to follow a similar trend to doping as the lattice orthorhombicity \cite{Prozorov}. Our data might thus reveal a pronounced sensitivity of the electronic properties to structural parameters, like the iron-pnictogen angle $\alpha$ \cite{calderon}, altered by external tunable variables like uniaxial pressure. Indeed, changes in $\alpha$ seem to induce relevant modifications in the shape of the Fermi surface and its nesting properties as well as in its orbital makeup, thus implying consequences in terms of the superconducting order parameter, critical temperature and magnetic properties \cite{calderon}.

The origin of the orthorhombic transition has been discussed from the closely related perspectives of spin fluctuations (a so called spin-induced nematic picture \cite{xu,johannes,fernandes,fradkin}), and also in terms of a more direct electronic effect involving, for instance, the orbital degree of freedom \cite{lv_anis,kruger,lee,lv,devereaux,valenzuela,bascones}. The present measurements alone cannot distinguish between these related scenarios, because in all cases some degree of electronic anisotropy is anticipated, and indeed it is likely that both orbital and spin degrees of freedom play a combined role in the real material. Nevertheless, it is instructive to compare the observed dichroism with specific predictions made within models based on orbital order. The two well-defined energy scales $\omega_1$ and $\omega_2$ (Fig. \ref{Sig1}) may represent optical transitions between states with the strongest $d_{xz}/d_{yz}$ character, which are separated by about 0.3-0.4 eV. Such an energy splitting is indeed compatible with the theoretical calculations of the anisotropic optical conductivity \cite{lv_anis} and of the linear dichroism in the X-ray absorption spectroscopy \cite{devereaux}. Nonetheless, the debate about the impact of orbital order on the electronic properties is far from being settle down. Valenzuela et al. point out that an increasing degree of the orbital order may favor a diminishing anisotropy of the excitation spectrum and above all of the Drude weights \cite{valenzuela}, opposite to what we may conclude from our optical experiment. Bascones et al. even claim that the orbital order accompanies the magnetization within a wide range of parameters but it is not correlated with the magnetic exchange anisotropy \cite{bascones}. While these latter calculations might be relevant at low temperatures, it remains to be seen how they can explain the onset of the anisotropy at/above $T_s$, where the spin symmetry is not yet broken. 
%
%
%
%
%
%
%
%
%
\begin{figure*}[ht]
\centering
\includegraphics[width=17cm]{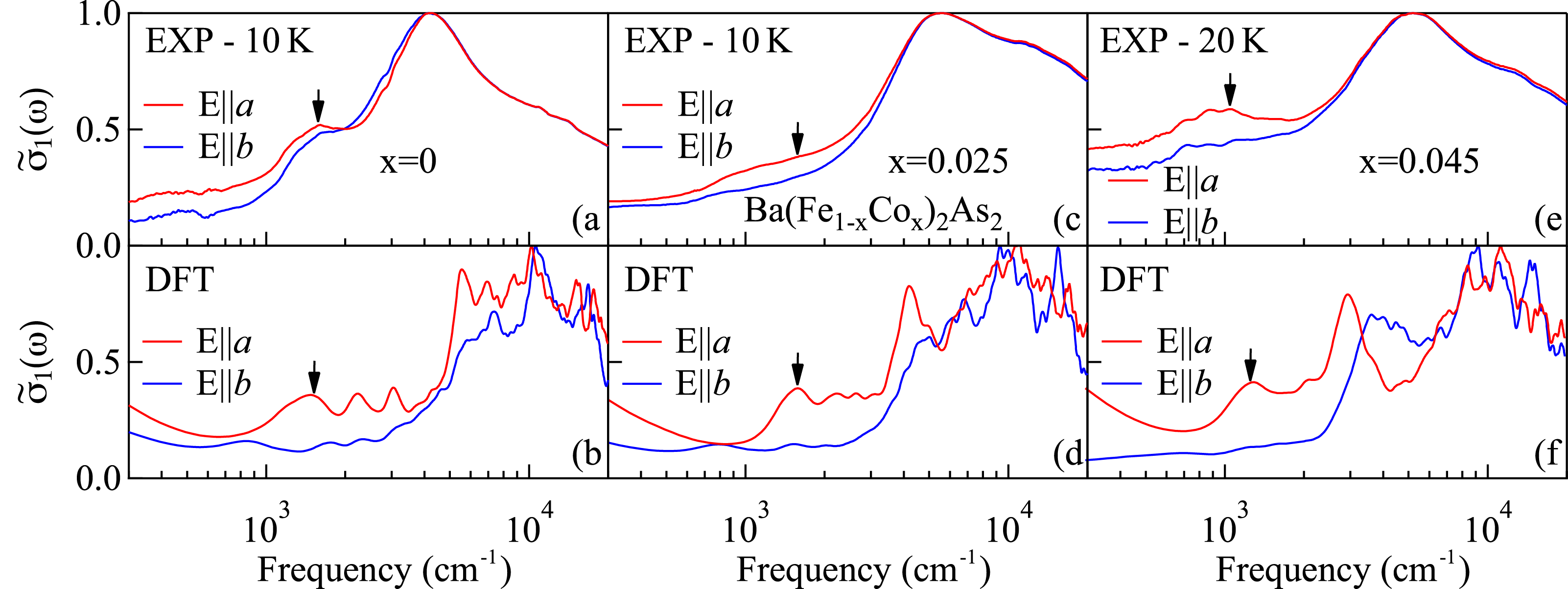}
\caption{(color online) Normalized optical conductivity $\tilde{\sigma}_{1}(\omega)$ of BaFe$_2$As$_2$ (a,b), Ba(Fe$_{0.975}$Co$_{0.025}$)$_2$As$_2$ (c,d), and Ba(Fe$_{0.955}$Co$_{0.045}$)$_2$As$_2$ (e,f) measured (top panels) at the lowest temperatures with light polarized along the $a$- and $b$-axis and obtained by DFT calculations (bottom panels) along the antiferromagnetically oriented spins and parallel to the ferromagnetic stripes \cite{sanna}. The black arrows indicate the MIR-band in the measurements and calculations.}\label{Exp_Theory}
\end{figure*}

Finally, we wish to come back to the comparison between our optical results and the outcome of the LAPW calculations \cite{sanna,software}. The model uses as initial lattice parameters those determined experimentally from the crystal structure of BaFe$_2$As$_2$ but neglects the small orthorhombic distortion occurring at low temperatures. The magnetic ordering is modeled using a fixed striped phase, where stripes of spins up are alternated with stripes of spins down on each Fe plane, compatible with the magnetic configuration determined experimentally below $T_N$ \cite{li}. The Co doping of  BaFe$_2$As$_2$ rapidly destabilizes the magnetic phase which disappears above 6\% Co concentrations. This Co-induced destabilization of magnetism was simulated by scaling the magnetic moment, calculated within the virtual crystal approximation \cite{sanna}, accordingly to the experimental critical temperature $T_N$ \cite{chu}. This approach corrects the deficiencies of the local spin density approximation and mimics the two main Co-induced effects: a localized electron doping on the Fe layer and a magnetic to nonmagnetic phase transition. The resulting theoretical Co-dopings agree within 0.5\% variation with the experimental ones.

Figure \ref{Exp_Theory} show the low temperature measured (top panels) and calculated (bottom panels) optical conductivity for both the $a$- and $b$-axis and for the three dopings. For a direct comparison we have normalized all the measured and calculated $\sigma_{1}(\omega)$ to their respective maxima, thus obtaining $\tilde{\sigma}_{1}(\omega)$. We clearly see a fairly good agreement between theory and experiment in the general shape of $\tilde{\sigma}_{1}(\omega)$. In the MIR range the DFT calculated $\tilde{\sigma}_{1}(\omega)$ finely reproduces the observed polarization dependence, both as far as the MIR-band position and its polarization dependence (black arrows in Fig. \ref{Exp_Theory}) are concerned. Indeed, the predictions of the enhancement of the MIR-band along the $a$-direction and its depletion along the $b$-direction are fairly close the experimental findings. Interestingly, the center of the calculated band (black arrows in Fig. \ref{Exp_Theory} bottom panels) shifts at lower frequencies for increasing dopings which is also in agreement with the experiments. This MIR-band is linked to the modeled magnetic stripe configuration which was shown to correspond to the energy-minimum configuration of these systems \cite{sanna}. Therefore, the DFT calculation strongly supports a "magnetic origin" of the MIR-band which would originate from the Fermi topology reconstruction in the magnetically ordered state. In this scenario one would reasonably expect that the MIR-band disappears above the magnetic phase transition, contrary to our observations. However, a dynamic antiferromagnetic order due to spin fluctuations could persist in the paramagnetic phase well above the phase transition temperature \cite{mazin}. The fingerprints of such underlying spin fluctuations would be frozen to fast enough probes as optics, thus explaining the persistence above the phase transition of the MIR-band in our spectra. At higher (NIR) frequencies the agreement result depleted because of the finite $k$-point sampling. The major interband peak of $\tilde{\sigma}_{1}(\omega)$ experimentally observed at about 5000 cm$^{-1}$ is shifted to slightly higher frequencies in the theoretical calculations and shows a steeper rise. We notice that spin-polarized DFT calculations require a smaller renormalization factor with respect to unpolarized ones in order to account for these frequency-shifts.

%
%
%
%
%
\section{Conclusions}
The charge dynamics of detwinned Ba(Fe$_{1-x}$Co$_x$)$_2$As$_2$ single crystals in the underdoped regime reveals an in-plane temperature and doping dependent optical anisotropy. At low frequencies the optical measurements offer the unique opportunity to disentangle the distinct behaviors of the Drude weights and scattering rates of the itinerant charge carriers, which are both enhanced along the antiferromagnetic $a$-axis with respect to the ferromagnetic $b$-axis. Our findings on such single domain specimens allow us to shed light on the counterintuitive anisotropic behavior ($\rho_b>\rho_a$) of the $dc$ resistivity. The $dc$ anisotropy below $T_N$ is principally determined by the anisotropy in the low frequency Drude weight (i.e., changes in the electronic structure close to the Fermi energy), outweighing the non-negligible anisotropy of the scattering rates between the $a$- and $b$-axis. Of equal or perhaps greater interest is the temperature regime above $T_N$ for which the Fermi surface is not reconstructed. One would like to understand whether the resistivity anisotropy at high temperatures also originates from the Fermi surface, perhaps due to the difference in the orbital occupancy revealed by ARPES \cite{devereaux,lv_anis}, or from anisotropic scattering, perhaps associated with incipient spin fluctuations \cite{fernandes2}. The current optical data may be in partial agreement with both points of view and therefore do not permit a conclusive answer to this question, thus motivating further experiments in order to definitely address the origin of the anisotropy above $T_N$.

The optical anisotropy extends to relatively high frequencies and temperatures above the phase transitions for crystals held under uniaxial stress. The resulting linear dichroism reveals the electronic nature of the structural transition and implies a substantial nematic susceptibility. In order to clarify the subtle interplay of magnetism and Fermi surface topology we elaborate on a comparison of our optical measurements with theoretical calculations obtained from density functional theory within the full-potential LAPW method. The calculations are able to reproduce most of the observed experimental features, in particular, to identify the MIR-band located at about 1500 cm$^{-1}$ as a magnetic peak, ascribed to antiferromagnetic ordered stripes. The measured large in-plane anisotropy of the optical response and its doping dependence is consistently tracked by the LAPW calculations. 
\\
\begin{acknowledgments}
The authors acknowledge fruitful discussions with S. Kivelson, T. Devereaux, C. Homes, D.N. Basov, R.M. Fernandes, J. Schmalian, W. Lv and D. Lu and valuable help by J. Johannsen in collecting part of the data. This work has been supported by the Swiss National Foundation for the Scientific Research within the NCCR MaNEP pool. This work is also supported by the
Department of Energy, Office of Basic Energy Sciences under
contract DE-AC02-76SF00515. The work in Cagliari is supported by the Italian MIUR through PRIN2008XWLWF.
\end{acknowledgments}

$^{*}$ Both authors equally contributed to the present work.

\end{document}